\documentclass[amsmath,amssymb,amsbsy,prb,twocolumn,superscriptaddress,showpacs]{revtex4-1}
\usepackage[dvipdfmx]{graphicx}
\usepackage{dcolumn}
\usepackage{bm}
\usepackage{braket}
\usepackage{mathtools}
\usepackage[breaklinks,colorlinks=true,linkcolor=blue,urlcolor=blue,citecolor=blue]{hyperref}
\usepackage{color}

\begin{document}


\title{Generic model for the hyperkagome iridate in the local-moment regime}

\author{Tomonari Mizoguchi}
\email{mizoguchi@hosi.phys.s.u-tokyo.ac.jp}
\affiliation{Department of Physics, University of Tokyo,
 7-3-1 Hongo, Bunkyo-ku, Tokyo 113-0033, Japan }
\affiliation{Department of Physics and Centre for Quantum Materials, University of Toronto, Toronto, Ontario M5S 1A7, Canada}

\author{Kyusung Hwang}
\affiliation{Department of Physics and Centre for Quantum Materials, University of Toronto, Toronto, Ontario M5S 1A7, Canada}

\author{Eric Kin-Ho Lee}
\affiliation{Department of Physics and Centre for Quantum Materials, University of Toronto, Toronto, Ontario M5S 1A7, Canada}

\author{Yong Baek Kim}
\affiliation{Department of Physics and Centre for Quantum Materials, University of Toronto, Toronto, Ontario M5S 1A7, Canada}
\affiliation{Canadian Institute for Advanced Research/Quantum Materials Program, Toronto, Ontario MSG 1Z8, Canada}
\affiliation{School of Physics, Korea Institute for Advanced Study, Seoul 130-722, Korea}

\begin{abstract}
The hyperkagome iridate, Na$_4$Ir$_3$O$_8$, has been regarded as a promising candidate material
for a three-dimensional quantum spin liquid. Here the three-dimensional network of 
corner-sharing triangles forms the hyperkagome lattice of Ir$^{4+}$ ions. Due to strong spin-orbit coupling,
the local moments of Ir$^{4+}$ ions are described by the pseudospin $j_{\rm eff} = 1/2$ Kramers doublet. 
The Heisenberg model on this lattice is highly 
frustrated and quantum/classical versions have been studied in earlier literature. In this work, we derive
a generic local-moment model beyond the Heisenberg limit for the hyperkagome iridate by considering
multi-orbital interactions for all the $t_{2g}$ orbitals and spin-orbit coupling. The lifting of massive classical
degeneracy in the Heisenberg model by various spin-anisotropy terms is investigated at the classical
level and the resulting phase diagram is presented.
We find that different anisotropy terms prefer distinct classes of magnetically ordered phases, often with
various discrete degeneracy. The implications of our results for recent $\mu$SR and NMR experiments 
on this material and possible quantum spin liquid phases are discussed.
\end{abstract}


\maketitle

\section{Introduction}

Recent budding interest on $5d$ transition metal oxides stems from the promise for emergent novel quantum
ground-states resulting from the cooperative effects of strong spin-orbit coupling and electron interactions.\cite{Witczak2014,Rau2015,Schaffer2015} 
For example, various topological phases are proposed to occur, which include topological insulator, topological 
semi-metal and quantum spin liquid phases as well as unusual magnetic states and superconductors.

In particular, iridates have enjoyed significant attention due to the availability of a variety of materials.\cite{Witczak2014,Rau2015,Schaffer2015} 
When spin-orbit coupling dominates over crystal field splitting, the basic electronic structure of
iridates with Ir$^{4+}$ ions can be described by the pseudospin $j_{\rm eff} =1/2$ Kramers doublet,\cite{Kim2008,Kim2009} which is
a combination of spin and orbital wave functions. Hence, in the strong-coupling limit, the Ir ions 
carry $j_{\rm eff} =1/2$ moments and the resulting interaction between them presents
highly quantum mechanical fluctuations. Such an interacting local moment system would be an ideal
platform for emergent quantum spin liquid states\cite{Balents2010} when it is placed on geometrically frustrated lattices
or the interaction itself has the capacity to generate massive classical degeneracy.

Currently two different classes of systems have been proposed for possible quantum spin liquid
phases in iridates. In the hyperkagome iridate, Na$_4$Ir$_3$O$_8$,\cite{Okamoto2007,Singh2013,Dally2014,Katayama2014,Fauque2015,Balodhi2015,Shockley2015} 
the Ir ions form a three-dimensional
network of corner-sharing triangles, which provides the geometric frustration for
the Heisenberg model. Indeed no magnetic ordering has been observed down to a few Kelvin
in spite of the large Curie-Weiss temperature, $\Theta_{\rm CW} \approx - 650K$.\cite{Okamoto2007} 
Possible quantum spin liquid and other related phases have been investigated via a variety of theoretical approaches.\cite{Hopkinson2007,Lawler2008,Lawler2008_2nd,Zhou2008,Chen2008,Podolsky2009,Motome2009,Norman2010,Norman2010_2nd,Bergholtz2010,Podolsky2011,Singh2012,Chen2013,Kimchi2014,Shindou2016}
In an alternative avenue, 2D (Na$_2$IrO$_3$, $\alpha$-Li$_2$IrO$_3$) honeycomb\cite{Singh2010,Liu2011,Ye2012} and 
3D ($\beta$- or $\gamma$-Li$_2$IrO$_3$) hyperhoneycomb\cite{Takayama2015,Modic2014}
iridates have been investigated 
in the context of bond-dependent interactions, such as the Kitaev interaction, 
between local moments\cite{Jackeli2009,Rau2014,Yamaji2014,Perkins2014,Lee2015,Kimchi2015}.
Such interactions, even if they are placed on bipartite lattices such as the honeycomb lattice, would 
generate extensive classical degeneracy and may lead to quantum spin liquid phases, as is the case for
the pure Kitaev model.\cite{Kitaev2006,Mandal2009,Schaffer_Kitaev2015,Motome2014,Trebst2016}

In this work, we investigate a generic local-moment model for the hyperkagome iridate, and we examine
the effects of anisotropic interactions beyond the Heisenberg limit. Earlier works on the hyperkagome
iridate were focused mostly on the Heisenberg model\cite{Hopkinson2007,Lawler2008,Lawler2008_2nd,Zhou2008,Bergholtz2010,Singh2012}
or the effects of selected sets of anisotropic interactions.\cite{Chen2008,Norman2010,Norman2010_2nd,Kimchi2014,Shindou2016}
Here we provide the derivation of a generic model for $j_{\rm eff}=1/2$ local moments of Ir$^{4+}$ ions 
by taking into account multiorbital interactions 
such as Hund's coupling
and spin-orbit coupling. 
In fact, effects of Hund's coupling were not considered in previous studies even though it is crucial for the appearance of anisotropic spin interactions in the hyperkagome iridate that has an edge-sharing structure of IrO$_6$ octahedra.\cite{Jackeli2009,Rau2014} The missing Hund's coupling is taken into account in our microscopic construction of the spin model.
Remarkably, various ``frustrating" bond-anisotropic interactions arise. The resulting model is characterized
by four parameters; $J$ for the antiferromagnetic Heisenberg exchange, $D$ for the Dzyaloshinskii-Moriya interaction,
$K$ for a Kitaev-like term, and $\Gamma$ for the symmetric anisotropic exchange.
The relative sign and form of the three different anisotropic exchange interactions depend on the bond directions and are
completely fixed by the lattice symmetries. 
As a result, the generic model for the hyperkagome iridate has both of the ingredients for massive classical 
degeneracy, namely geometric frustration and frustrated exchange interactions.
Indeed all of the $J$-only, $K$-only, $\Gamma$-only models are frustrated and support distinct sets
of a classically-degenerate manifold. This is in contrast to the honeycomb or hyperhoneycomb systems,
in which $J$-only model leads to a unique antiferromagnetically-ordered ground-state.

Given that the Curie-Weiss temperature is large and negative, we assume that $J \sim 300K$ is the dominant
energy scale and examine the effects of small anisotropic interactions represented by $D$, $K$, and $\Gamma$.
This approach is motivated by recent $\mu$SR and NMR experiments in which short-range spin correlations
and/or some kind of spin freezing behaviors have been discovered below $T \sim 6$-$7K$.\cite{Dally2014,Shockley2015}
The idea is that various small anisotropic interactions become much more important below 6-7$K$ and the 
understanding of the nature of the ground-state may require careful examination of the effects of
these small perturbations, while the physics of the higher temperature phase may still be understood using the 
Heisenberg interaction.

\begin{figure*}
\centering
\includegraphics[trim= 0mm 40mm 0mm 0mm, clip, width=\linewidth]{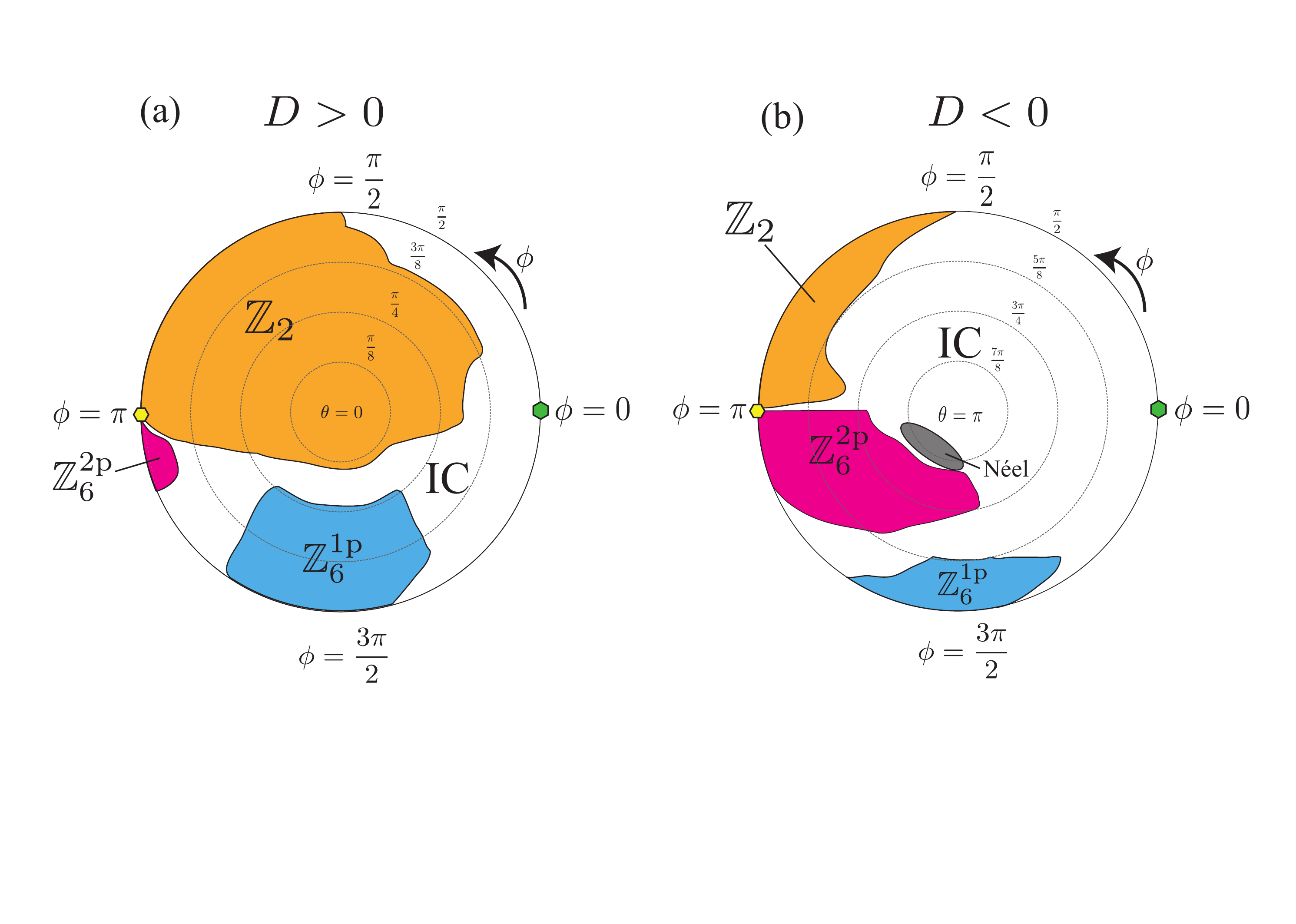}
\caption{
(Color online) Phase diagrams of the generic $J$-$K$-$\Gamma$-$D$ model in Eq. (\ref{eq:Hamiltonian}).
In the diagrams, the coupling constants are parametrized in terms of the two variables, $\theta$ and $\phi$, as defined in Eq. (\ref{eq:parametrization}).
The left and right diagrams represent the cases with (a) $D>0$ and (b) $D<0$, respectively.
In each case, the center and the circumference of the disk diagram represent the $J$-$D$ ($\theta=0$ or $\pi$) and 
$J$-$K$-$\Gamma$ ($\theta=\pi/2$) models, respectively.
These two limits are interpolated by moving along the radial ($\theta$) and/or circumferential ($\phi$) directions.
The diagrams highlight three $q=0$ noncoplanar magnetic phases: $\mathbb{Z}_2$ windmill (orange), $\mathbb{Z}_6^{2p}$ (pink), and $\mathbb{Z}_6^{1p}$ (light blue).
These three  $q=0$ phases compete with the incommensurate phase (white).
The diagrams contain other magnetic phases such as the N\'eel (gray) and ferrimagnetic (green) phases.
The yellow dot indicates a special point where the $\mathbb{Z}_2$ windmill and $\mathbb{Z}_6^{2p}$ states become degenerate and form together the ground-state manifold.
The $q=0$ noncoplanar phases are described in Sec. \ref{sec:q=0order}.
} 
\label{fig:JKDG}
\end{figure*}

As the first step towards this goal, we map out the magnetic phase diagram for the classical model using the 
Luttinger-Tisza\cite{Luttinger1946,Luttinger1951} and classical Monte Carlo simulated annealing methods.
In particular, we examine how the massive classical degeneracy of the Heisenberg model is lifted depending on which anisotropic 
interaction is dominant. The resulting phase diagram for a selected set of parameters is shown in Fig. \ref{fig:JKDG}.
There exist three dominant $q=0$ ground-state manifolds with $\mathbb{Z}_2$ or two distinct $\mathbb{Z}_6$ discrete degeneracies,
labelled as $\mathbb{Z}^{1p}_6$ and $\mathbb{Z}^{2p}_6$, in addition to an incommensurate magnetic order. 
Here $\mathbb{Z}^{1p}_6$ ($\mathbb{Z}^{2p}_6$) refers to the manifold of classical ground-states where the direction of one (two) of the moments
on each triangle is almost parallel to the local $C_2$ axis at each site,
which we will define later,
while the remaining two (one) are not (see Figs. \ref{fig:Z6_2p} and \ref{fig:Z6_1p}).
When $D > 0$ and $\Gamma > 0$, the $q=0$ state with $\mathbb{Z}_2$ degeneracy (which is called
the canted windmill state\cite{Chen2008}) is the ground-state. If $D > 0$ and $\Gamma < 0$, the $q=0$ state with
$\mathbb{Z}^{1p}_6$ degeneracy is dominant for relatively large $\Gamma$.
On the other hand, when $D < 0, K < 0,$ and $\Gamma < 0$, the $q=0$ state with $\mathbb{Z}^{2p}_6$ degeneracy
becomes dominant for large $K$. In general, $D, \Gamma, K$ promote the $q=0$ states with 
$\mathbb{Z}_2, \mathbb{Z}^{1p}_6, \mathbb{Z}^{2p}_6$ degeneracy, respectively (see the phase diagrams in Figs. \ref{fig:windmill}, \ref{fig:Z6_2p}, and \ref{fig:Z6_1p}).

If the $q=0$ states with discrete degeneracy dominate the low temperature short-range spin correlations,
one of the degenerate states, once it is formed locally in certain regions, may not easily relax to another
degenerate configurations. This is due to the constrained spin dynamics in frustrated magnets, 
as explained in the main text.
The energy/temperature scale where these phenomena occur will be set by the dominant
anisotropic interactions. This may explain the spin freezing or slow spin dynamics observed in the experiments.
Our results also suggest that it would be fruitful to investigate quantum spin liquid phases that
may be obtained by quantum disordering the $q=0$ states described above.

\section{Hyperkagome lattice\label{sec:lattice}}

\begin{figure}
\centering
\includegraphics[width=\linewidth]{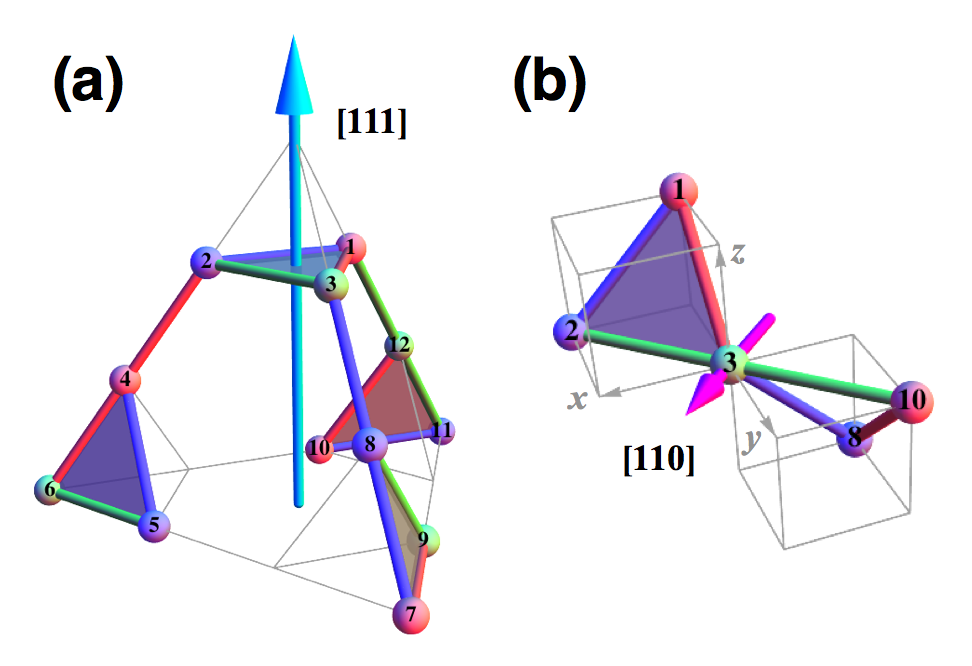}
\vspace{-10pt}
\caption{
(Color online) Symmetries of the hyperkagome lattice.
(a) The global $C_3$ rotation.
Corresponding to each triangle in the hyperkagome lattice, there exists global $C_3$ rotation symmetry.
The cyan arrow along the [111] direction represents the $C_3$ axis at the triangle formed by the sites 1, 2, 3. 
(b) The local $C_2$ rotation.
For each site on the hyperkagome lattice, there are two triangles sharing the site.
The two triangles are related by a local $C_2$ rotation.
The pink arrow shows the local $C_2$ axis for the site 3, which is along the [110] direction.
} 
\label{fig:symm}
\end{figure}

\begin{table}
\centering
\caption{
Site classification and local $C_2$ axes.
The table lists the site type and $C_2$ axis for the 12 sites in a unit cell (described in Fig. \ref{fig:lattice}).
\label{tab:localc2}
}
\begin{ruledtabular}
\begin{tabular}{ccc|ccc}
Site &Type & $C_2$ axis & Site & Type  & $C_2$ axis
\\
\hline
1 & $x$ & $[011]$ & 7 & $x$ &$[011]$
\\
2 & $y$ & $[101]$ & 8 & $y$ & $[10\bar{1}]$
\\
3 & $z$ & $[110]$ & 9 & $z$ &$[1\bar{1}0]$
\\
\hline
4 & $x$ & $[01\bar{1}]$  & 10 & $x$ &$[01\bar{1}]$
\\
5 & $y$ & $[10\bar{1}]$ & 11 & $y$ &  $[101]$
\\
6 & $z$ & $[110]$& 12 & $z$ & $[1\bar{1}0]$
\\
\end{tabular} 
\end{ruledtabular}
\end{table}
We start with a brief introduction on the structure and symmetries of the hyperkagome lattice, which will be 
used to describe the local-moment model introduced in the next section.
\begin{figure}
\centering
\includegraphics[width=\linewidth]{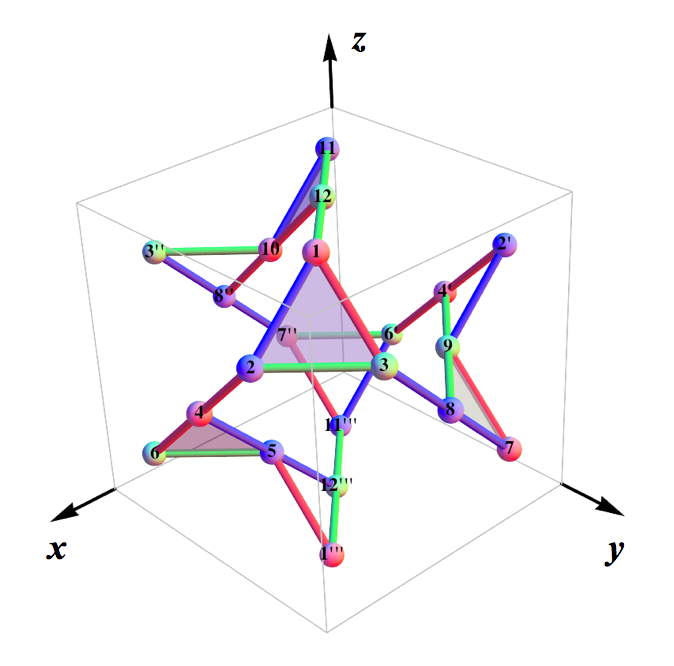}
\vspace{-10pt}
\caption{
(Color online) Structure of the hyperkagome lattice.
The figure shows the 12 sites (labelled with $1,\cdots,12$) and 24 nearest-neighbor bonds in a cubic unit cell.
The sites with a primed number mean sites that belong to a neighboring unit cell. 
The sites (nearest-neighbor bonds) in red, blue, green represent the $x$-, $y$-, $z$-sites (bonds), respectively.
For the classifications of the sites and bonds, see Sec. \ref{sec:lattice}.
\label{fig:lattice}
} 
\end{figure}
The hyperkagome lattice is a three dimensional network of corner-sharing triangles, and it can be thought of
as a higher dimensional version of the kagome lattice.
In contrast to the kagome lattice, however, the corner-sharing triangles are not coplanar and have
different orientations chosen from (111), ($\bar{1}$11), (1$\bar{1}$1), and (11$\bar{1}$) planes, 
leading to a large cubic unit cell with 12 sites/sublattices and 24 nearest-neighbor (NN) bonds (Fig. \ref{fig:lattice}).
The lattice is characterized by several symmetries that are useful to describe the model and the magnetic structure.
First, there exists the $C_3$ rotational symmetry with respect to the $C_3$ axis through the center of each triangle. 
For example, the [111] axis through the triangle formed by the sites 1, 2, and 3 represents such a rotation symmetry [Fig. \ref{fig:symm} (a)].
Another useful symmetry is the 
$C_2$ rotational symmetry with local $C_2$ axis defined at each site\cite{Chen2008}.
As described in Fig. \ref{fig:symm} (b), for each site 
there exists a $C_2$ rotation that transforms one triangle sharing the same site into the other.
Due to the $C_3$ and $C_2$ rotational symmetries, all the NN bonds are equivalent on the hyperkagome lattice.

We find it useful to classify the NN bonds on the hyperkagome lattice into three categories.
The NN bonds are labeled as the $x$-, $y$-, and $z$-bonds if they are parallel to the $yz$, $zx$, and $xy$ planes in the global coordinates, respectively.
The $x$-, $y$-, and $z$-bonds are denoted by red, blue, and green, respectively, in Figs. \ref{fig:lattice} and \ref{fig:symm}.
This bond classification leads to a natural characterization of the sites/sublattices.
Each site has four NN bonds, two of which make a straight line through the site and hence are parallel to each other.
We now label each site in terms of the bond-type of the two parallel bonds.
For instance, if there are two parallel $z$-bonds for a given site, this site is labeled as the $z$-site [see site 3 in Fig. \ref{fig:symm} (b)].
We use the same color scheme for the sites as for the bonds, {\it i.e.}, the $x$-, $y$-, $z$-sites are denoted by red, blue, green, respectively.

The aforementioned $C_2$ axis at a site is closely related to the character of the site. 
For example, the $C_2$ axis at a $z$-site is perpendicular to the $z$ axis and also the bond direction defined by the two parallel NN bonds for the site.
Accordingly, site 3 ($z$-site) with the [1$\bar{1}$0] bond direction has the [110] $C_2$ axis [Fig. \ref{fig:symm} (b)].
The local $C_2$ axes for the 12 sublattices as well as the site classification are summarized in Table \ref{tab:localc2}.

\section{Model}
Now we introduce a generic symmetry-allowed Hamiltonian for the hyperkagome iridate Na$_4$Ir$_3$O$_8$ in terms of the
$j_{\rm eff}=1/2$ moment represented by $\bm{S}$.\cite{Chen2008}
The Hamiltonian consists of the isotropic Heisenberg interaction ($J$) and three different anisotropic interactions: bond-dependent Kitaev ($K$), 
Dzyaloshinskii-Moriya ($D$), and anisotropic and symmetric (${\Gamma}$) interactions.
\begin{align}
\mathcal{H}&=
\sum_{\langle ij \rangle \in \alpha}
\left [ 
J \bm{S}_{i} \cdot \bm{S}_{j} 
+  K S^{\alpha}_{i} S^{\alpha}_{j} \right ] \notag \\
&+  
\sum_{\langle ij \rangle \in \alpha, \beta \gamma}
\left [ D  \eta_{ij} ( S^{\beta}_{i} S^{\gamma}_{j} - S^{\gamma}_{i} S^{\beta}_{j} )
+   \Gamma \xi_{ij} ( S^{\beta}_{i} S^{\gamma}_{j} + S^{\gamma}_{i} S^{\beta}_{j} ) \right ] .
\label{eq:Hamiltonian}
\end{align}
Here the Kitaev term $K$ represents the bond-dependent Ising interaction $S^{\alpha}_{i} S^{\alpha}_{j}$ for an $\alpha$-type
NN bond $ij$ or $\langle ij \rangle \in \alpha$ (where $\alpha=x,y,z$).
In the Dzyaloshinskii-Moriya $D$ and anisotropic and symmetric $\Gamma$ interactions, the shorthand notation $\langle ij \rangle \in \alpha, \beta \gamma$ 
means that for an $\alpha$-type bond $ij$, $\beta$ and $\gamma$ are fixed in such a way that $\alpha\beta\gamma$ is a cyclic permutation of $xyz$.
The bond-dependent sign factors, $\eta_{ij}$, $\xi_{ij}~(=\pm)$, determined by the $C_3$ and $C_2$ symmetries mentioned above
are summarized in Table \ref{tab:bond_sign}.
Throughout the paper, we fix the Heisenberg coupling to be 1 ($J=1$).

The Hamiltonian described above can be derived explicitly by using a strong-coupling expansion for the $t_{2g}$ electrons with the Kanamori-type multiorbital interactions
such as the Coulomb interaction $U$ and Hund's coupling $J_H$.\cite{Kanamori1963,Chen2013,Sugano1970,Rau2014} 
Here the strong spin-orbit coupling is taken into account via the projection of 
the $t_{2g}$ manifold into the $j_{\textup{eff}}=1/2$ Kramers doublet after the large interaction limit is taken.
In this work, we consider an ideal structure for the edge-sharing IrO$_6$ octahedra of the hyperkagome iridate, and we use the Slater-Koster parametrization\cite{Slater1954} to 
represent the hopping amplitudes between $t_{2g}$ orbitals. 
Details of this derivation are provided in Appendix \ref{sec:modelderivation}.

In the rest of this paper, we investigate the classical ground-states of this model Hamiltonian.
For convenience, we will often refer to the $j_{\rm eff} = 1/2$ moment as ``spin" in the discussion of the classical ground-states.
First, we notice that the ground-states are highly degenerate in the Heisenberg limit ($K=D={\Gamma}=0$).
The degenerate ground-states are characterized by 120$^{\circ}$ spin structure on each triangle, {\it i.e.}, ${\bf S}_i+{\bf S}_j+{\bf S}_k=0$ 
among the three spins on the triangle.
When the Heisenberg model is perturbed by anisotropic interactions as assumed for the hyperkagome iridate, the macroscopic ground-state degeneracy will be lifted.
To investigate the ground-states selected by the anisotropies and their physical properties,
we solve the classical model by employing the Luttinger-Tisza and simulated annealing methods.
The resulting phase diagram is shown in Fig. \ref{fig:JKDG}.
In the next section, we explain three major $q=0$ noncoplanar magnetically-ordered states 
shown in this phase diagram.

\begin{table}
\centering
\caption{
Bond classification and the bond-dependent sign factors, $\eta_{ij}$ and $\xi_{ij}$, of the Hamiltonian $\mathcal{H}$.
The table lists the bond character ($\alpha$) and the sign factors ($\eta_{ij}$, $\xi_{ij}$) for the 24 nearest-neighbor bonds in a unit cell (depicted in Fig. \ref{fig:lattice}).
\label{tab:bond_sign}
}
\begin{ruledtabular}
\begin{tabular}{ccc|ccc}  
$( i,j )_\alpha$ & $\eta_{ij}$ & $\xi_{ij}$ & $( i,j )_\alpha$ & $\eta_{ij}$ & $\xi_{ij}$
\\
\hline
(1,2)$_y$ & $-$ & $-$ & $(7'',11''')_x$ & $-$ & $-$ 
\\
(2,3)$_z$ & $-$ & $-$ & $(11''',6')_y$ & $-$ & $-$
\\
(3,1)$_x$ & $-$ & $-$ & $(6',7'')_z$ & $-$ & $-$
\\
&&&&&
\\
(4,5)$_y$ & $+$ & $+$ & $(1''',5)_x$ & $-$ & $-$
\\
(5,6)$_z$ & $-$ & $-$ & $(5,12''')_y$ & $+$ & $+$
\\
(6,4)$_x$ & $+$ & $+$ & $(12''',1''')_z$ & $+$ & $+$
\\
&&&&&
\\
(7,8)$_y$ & $+$ & $+$ & $(10,8'')_x$ & $+$ & $+$
\\
(8,9)$_z$ & $+$ & $+$ & $(8'',3'')_y$ & $+$ & $+$
\\
(9,7)$_x$ & $-$ & $-$ & $(3'',10)_z$ & $-$ & $-$
\\
&&&&&
\\
(10,11)$_y$ & $-$ & $-$ & $(4',2')_x$ & $+$ & $+$
\\
(11,12)$_z$ & $+$ & $+$ & $(2',9)_y$ & $-$ & $-$
\\
(12,10)$_x$ & $+$ & $+$ & $(9,4')_z$ & $+$ & $+$
\\
\end{tabular}
\end{ruledtabular}
\end{table}

\section{q $=$ 0 noncoplanar states\label{sec:q=0order}}
In our analysis for the classical limit of the Hamiltonian, we find three $q=0$ noncoplanar states that take fairly large regions in the parameter space.
The three states can be represented as (i) $\mathbb{Z}_2$ windmill, (ii) $\mathbb{Z}_6^{2p}$, and (iii) $\mathbb{Z}_6^{1p}$ states and 
they are labeled by the discrete degeneracy (subscript) and the character of the spin configurations (superscript).
The discrete degeneracy of each state can be understood from the time-reversal and/or $C_3$ rotation symmetries as explained below.

\subsection{$\mathbb{Z}_2$ windmill states}
\begin{figure}
\centering
\includegraphics[width=0.9\linewidth]{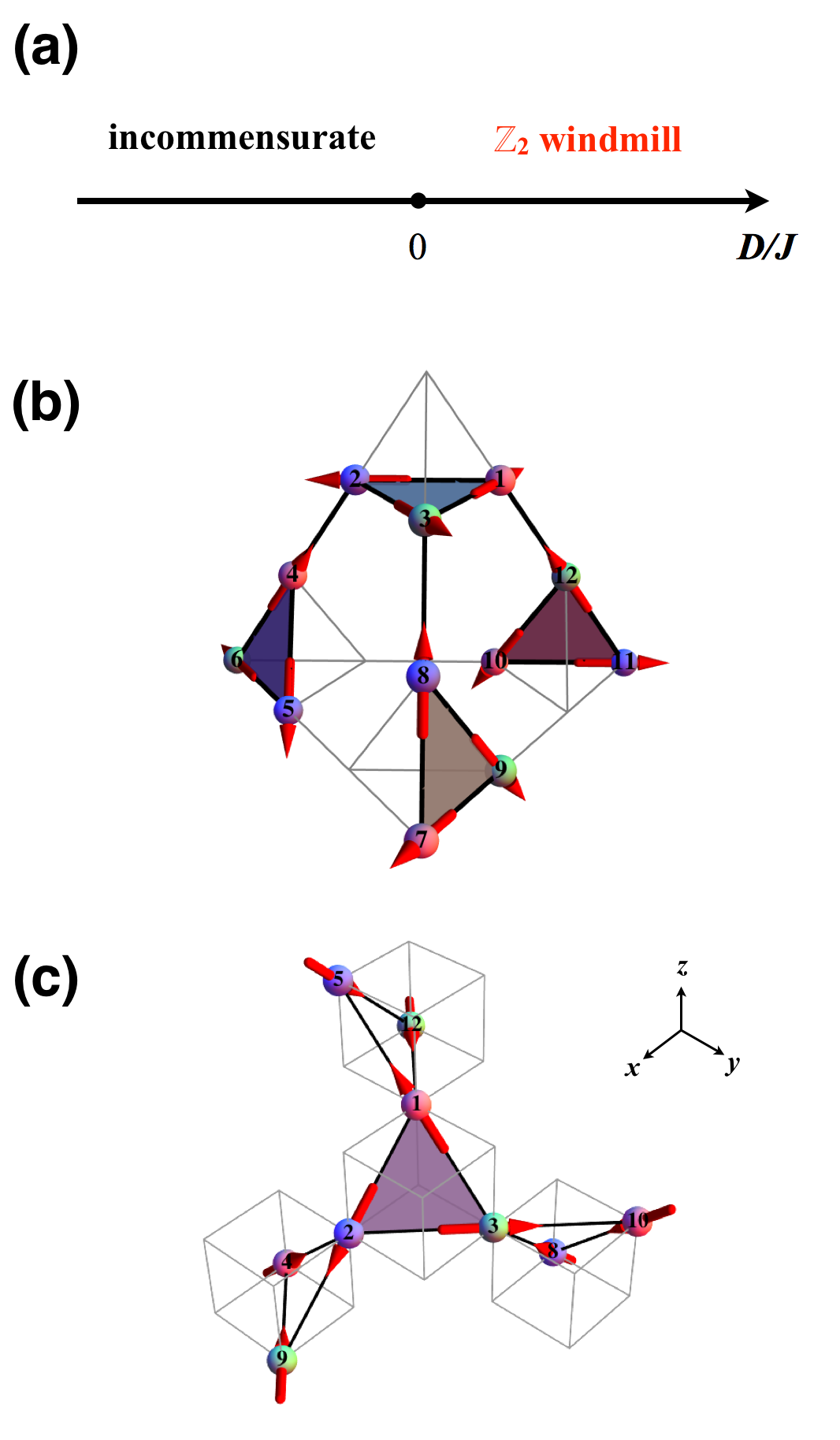}
\caption{
(Color online) Phase diagram of the $J$-$D$ model and visualization of the $\mathbb{Z}_2$ windmill states.
(a) In the $J$-$D$ model, the $\mathbb{Z}_2$ windmill states appear as the ground-states when $D>0$.
(b, c) Spin configuration of one of the $\mathbb{Z}_2$ windmill states in the (b) pyrochlore frame and (c) cubic frame.
For simplicity, an ideal windmill state with no canting is presented.
Spin moments (red arrows) in the windmill state point in the local bond directions, forming the 120$^{\circ}$ spin structure on each triangle.
}
\label{fig:windmill} 
\end{figure}

The $\mathbb{Z}_2$ (doubly degenerate) windmill states are featured with the 120$^{\circ}$ spin structures where 
the spin moment at each site is aligned along the local bond direction defined by the direction of two
parallel NN bonds or the straight line formed by two NN bonds sharing the given site.
Hence, there are two possible choices for the direction of the spin moment at each site.
Once the direction is chosen for one of the sites, and if we arrange the spin moments at other sites
to satisfy the 120$^{\circ}$ spin structure at every local triangle, we obtain one of the $\mathbb{Z}_2$ windmill states.
The other windmill state is obtained by acting the time reversal on the former state.
One of the windmill state is shown in Fig. \ref{fig:windmill}.
Notice that the windmill states are invariant under all of the $C_3$ rotations about
the [111], [$\bar{1}$11], [1$\bar{1}$1], and [11$\bar{1}$] axes. 

The ideal windmill structure described above occurs only at some special places in the phase diagram.
In general, the spin moments are slightly canted out of the local triangular planes, but the overall spin structure
still preserves the $C_3$ rotation invariance and twofold degeneracy given by the time reversal.
The canting is attributed to the effect of the anisotropic spin interactions.
The simplest model that allows the windmill states is the $J$-$D$ model.
When the Heisenberg model is perturbed with the positive Dzyaloshinskii-Moriya (DM) interaction ($D>0$), the canted windmill states 
with the twofold degeneracy appear as the ground-states (Fig. \ref{fig:windmill}).
The net canting component at a local triangle is perpendicular to the triangular plane.
With the negative DM interaction ($D < 0$) for the $J$-$D$ model, we find incommensurate states in the ground-state manifold.
The canted windmill states were discussed in Ref. [\onlinecite{Chen2008}] as the classical ground-states selected by the Dzyaloshinskii-Moriya interaction.

\subsection{$\mathbb{Z}_6^{2p}$ states}

\begin{figure}
\centering
\includegraphics[width=0.9\linewidth]{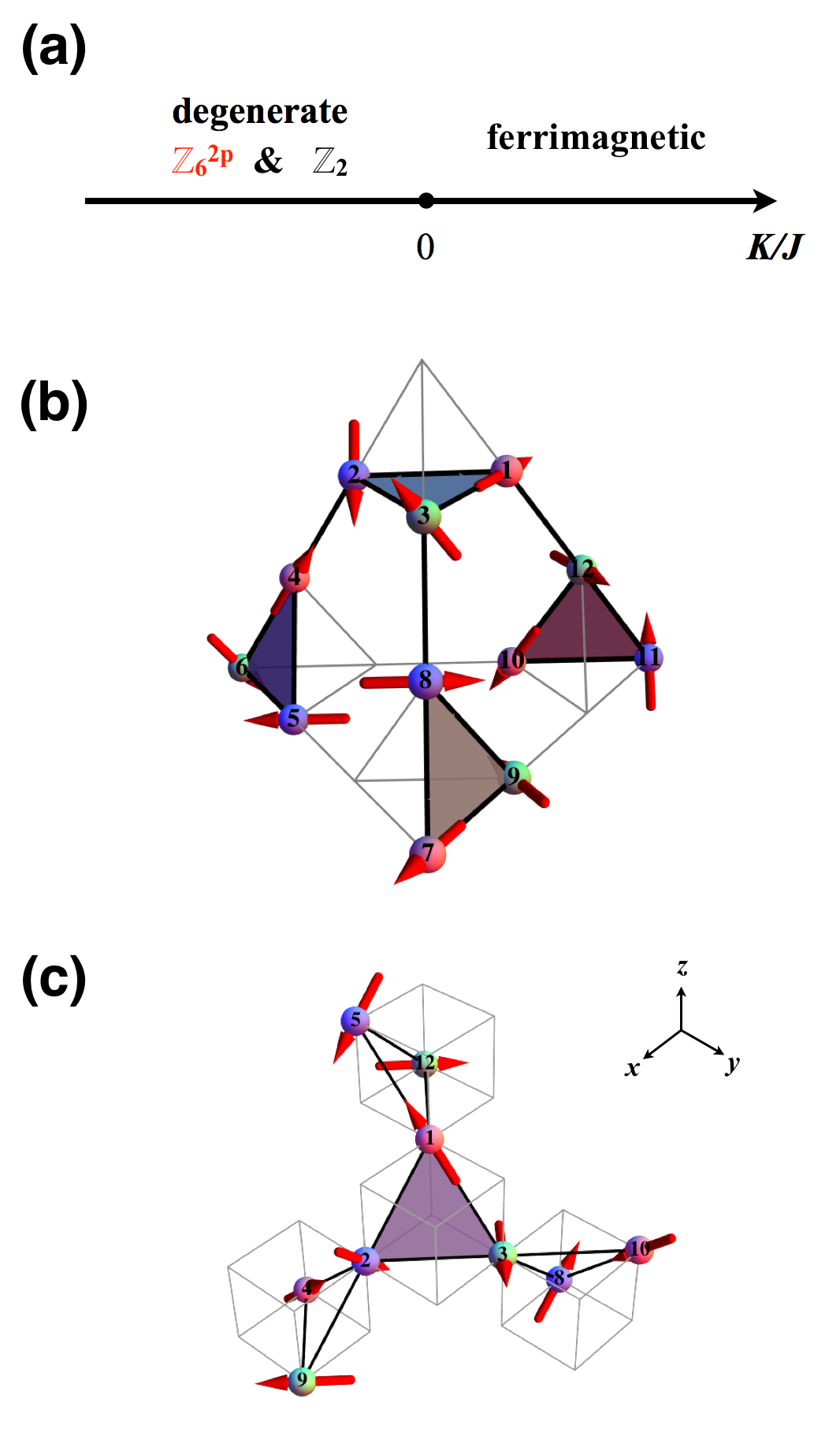}
\caption{
(Color online) Phase diagram of the $J$-$K$ model and visualization of the $\mathbb{Z}_6^{2p}$ states.
(a) When $K<0$ in the $J$-$K$ model, the $\mathbb{Z}_6^{2p}$ states form the ground-state manifold together with the $\mathbb{Z}_2$ windmill states.
(b, c) Spin configuration of one of the $\mathbb{Z}_6^{2p}$-$yz$ states in the (b) pyrochlore frame and (c) cubic frame.
For simplicity, an idealized $\mathbb{Z}_6^{2p}$-$yz$ state with no canting is presented.
In the $\mathbb{Z}_6^{2p}$-$yz$ state,
spin moments at the $y$- and $z$-sites (blue and green) are parallel to the local $C_2$ axes while moments at the $x$-sites (red) aligned along the local bond directions.
\label{fig:Z6_2p}
} 
\end{figure}

In the spin configurations of the $\mathbb{Z}_6^{2p}$ states, only two out of the three types of sites ($x$-, $y$-, and $z$-sites) have spin moments
parallel to the local $C_2$ axes. This explains the superscript $2p$ in $\mathbb{Z}_6^{2p}$.
Let us first consider three states that will be called $yz$, $zx$, $xy$ states.
For example, the $yz$ state means that the spins at the $y$- and $z$-sites are parallel to the local $C_2$ axes,
respectively. In contrast, the spins at the $x$-sites are perpendicular to the local axes, and at the same time parallel to the bond directions.
One of the $yz$-type states is visualized in Fig. \ref{fig:Z6_2p}. 
As seen in the figure, only the spin moments at the $x$-sites are parallel to the local triangular planes.
The $zx$ and $xy$ states are defined in similar ways. 
Notice that $yz$, $zx$, $xy$ states are related to each other by the $C_3$ rotations. 
One can now obtain the other three states by acting the time reversal on the former three states.
The latter three states obtained in this way are also related to each other by the $C_3$ rotations.
Hence the six states in the $\mathbb{Z}_6^{2p}$ manifold can be divided into two groups,
each having three states related to each other by the $C_3$ rotations, and these two groups
are transformed to each other by the time reversal.

As in the windmill states, the $\mathbb{Z}_6^{2p}$ states generally have canting of the spin moments from the idealized spin configurations described above.
The $J$-$K$ model is the simplest model where one can find the $\mathbb{Z}_6^{2p}$ ground-states.
When the Kitaev interaction is ferromagnetic ($K<0$), the $\mathbb{Z}_6^{2p}$ states appear (as well as the $\mathbb{Z}_2$ windmill states) in the 
ground-state manifold (see Fig. \ref{fig:Z6_2p}). With the antiferromagnetic Kitaev interaction ($K>0$), we find ferrimagnetic ground-states with 
eightfold degeneracy.
Details of the Luttinger-Tisza analysis for the $J$-$K$ model are provided in Appendix \ref{sec:LTA-JK}.

\subsection{$\mathbb{Z}_6^{1p}$ states}
\begin{figure}
\centering
\includegraphics[width=0.9\linewidth]{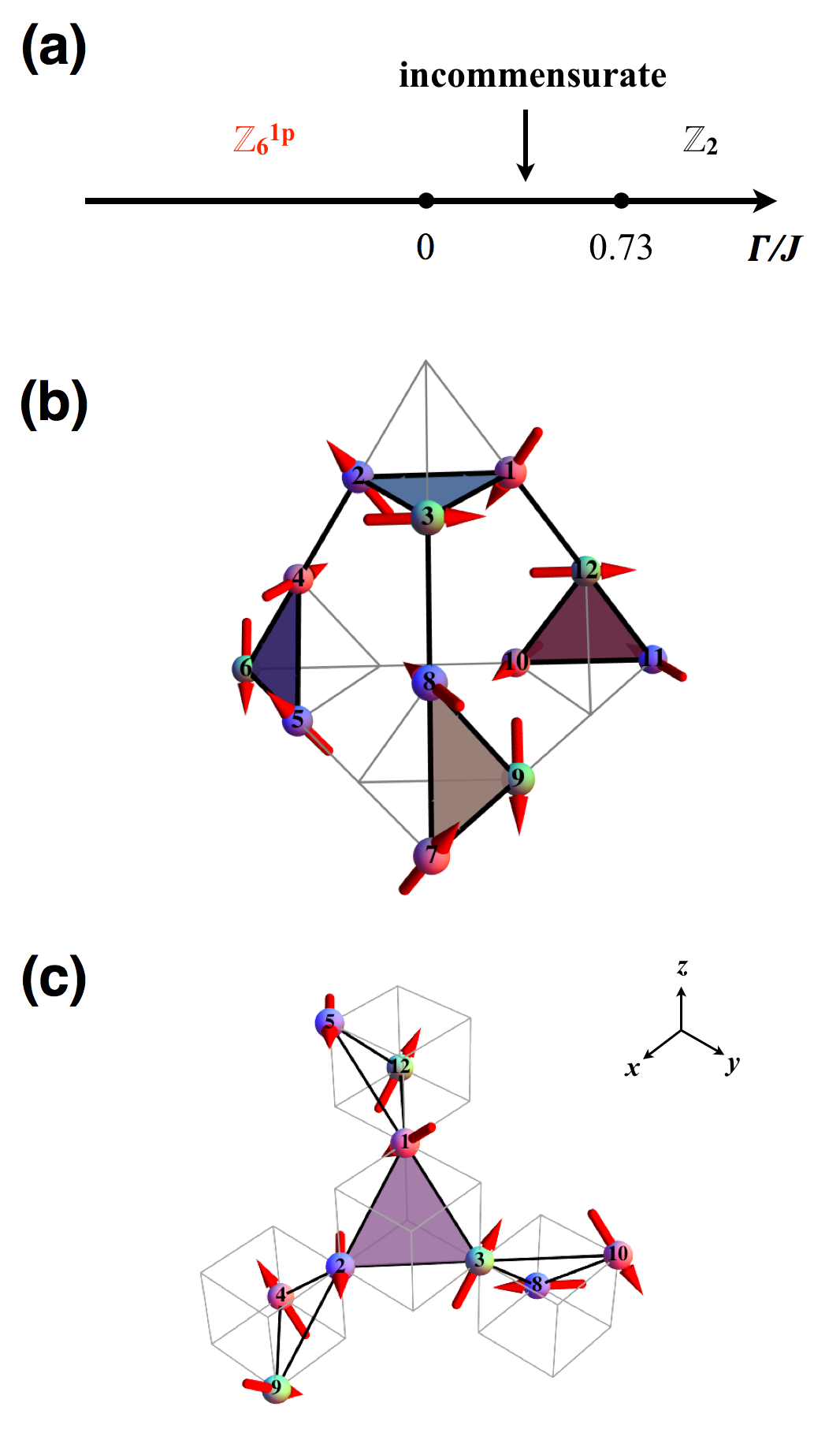}
\caption{
(Color online) Phase diagram of the $J$-$\Gamma$ model and visualization of the $\mathbb{Z}_6^{1p}$ states.
(a) In the $J$-$\Gamma$ model, the $\mathbb{Z}_6^{1p}$ ground-states are realized when $\Gamma<0$.
(b, c) Spin configuration of one of the $\mathbb{Z}_6^{1p}$-$x$ states in the (b) pyrochlore frame and (c) cubic frame.
In the $\mathbb{Z}_6^{1p}$-$x$ state,
spin moments at the $x$-sites (red) point along the local $C_2$ axes while moments at the $y$- and $z$-sites (blue and green) are lying along the $xy$- and $xz$-planes, respectively.
\label{fig:Z6_1p}
} 
\end{figure}

The $\mathbb{Z}_6^{1p}$ states can be characterized similarly to the $\mathbb{Z}_6^{2p}$ states.
The sixfold degeneracy and behaviors under the $C_3$ rotations as well as the time reversal that we discussed for the latter are also found in the former.
As implied by the superscript $1p$, an important difference between them is the number of types of sites that have spins along the local $C_2$ axes.
For the $\mathbb{Z}_6^{1p}$ states,
three states related by the $C_3$ rotations are labeled as the $x$, $y$, $z$ states.
In the $x$ state, only at the $x$-sites are the spin moments parallel to the local axes (see Fig. \ref{fig:Z6_1p}).
Interestingly, at the $y$- and $z$-sites, the spin moments are lying along the $xy$- and $xz$-planes.
This is another point that differentiates the $\mathbb{Z}_6^{1p}$ states from the $\mathbb{Z}_6^{2p}$ states.
The $y$ and $z$ states are defined similarly. Again the other three states can be obtained by
the time reversal.

The $J$-$\Gamma$ model provides the simplest setting for the $\mathbb{Z}_6^{1p}$ ground-states.
On the negative side of the anisotropic and symmetric interaction ($\Gamma<0$), the $\mathbb{Z}_6^{1p}$ ground-states are found with the ideal structure described above (Fig. \ref{fig:Z6_1p}).
However, the canting of the spin moments is generated in the $\mathbb{Z}_6^{1p}$ states when more than two anisotropic interactions exist as we shall see later.
In the other case with $\Gamma>0$, one can find incommensurate states and the windmill states and they are separated by the phase boundary $\Gamma/J=0.73$.
We provide details of the Luttinger-Tisza analysis for the $J$-$\Gamma$ model in Appendix \ref{sec:LTA-JG}.

We also provide static spin structure factors in Appendix \ref{sec:SSF} to further characterize the above $q=0$ states ($\mathbb{Z}_2$, $\mathbb{Z}_6^{2p}$, $\mathbb{Z}_6^{1p}$).

\section{Interplay of two different anisotropies}
\begin{figure*}
\centering
\includegraphics[width=\linewidth]{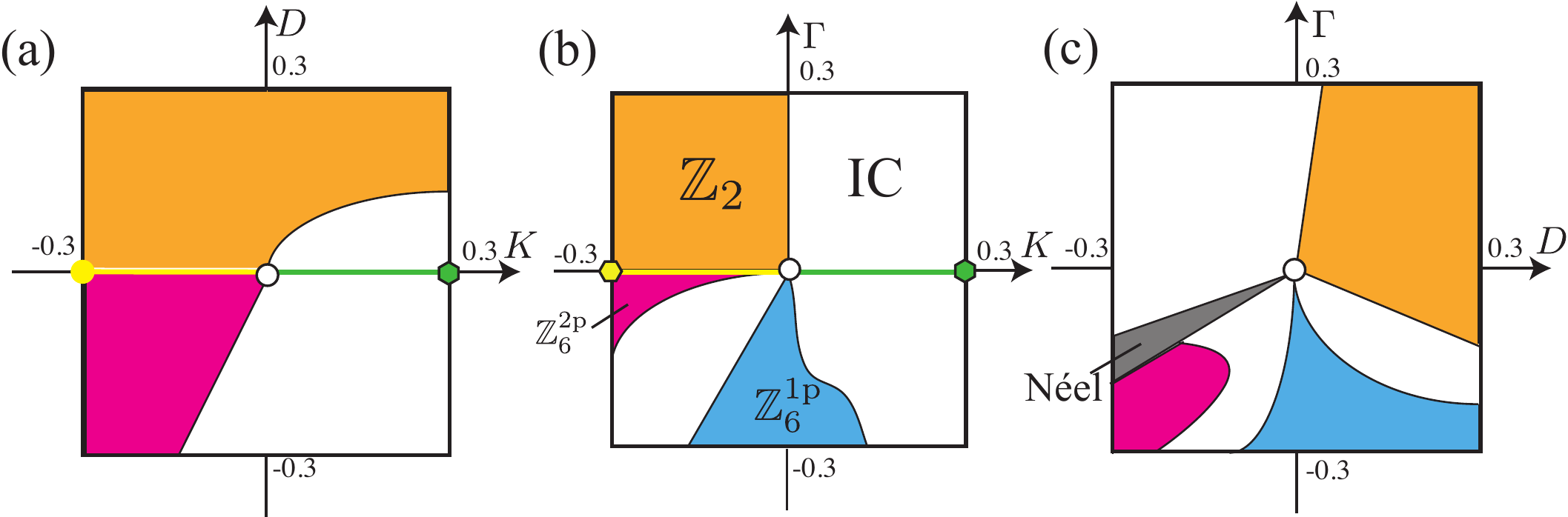}
\caption{
(Color online) Phase diagrams of the (a) $J$-$K$-$D$, (b) $J$-$K$-$\Gamma$, and (c) $J$-$D$-$\Gamma$ models.
The phase diagrams show extension of the $q=0$ noncoplanar states by the interplay of two different anisotropies: $\mathbb{Z}_2$ windmill (orange), $\mathbb{Z}_6^{2p}$ (pink), and $\mathbb{Z}_6^{1p}$ (light blue).
The diagrams contain other magnetic phases such as the ferrimagnetic (green), N\'eel (gray), and incommensurate (white) phases.
The yellow line represents a special case where the $\mathbb{Z}_2$ and $\mathbb{Z}_6^{2p}$ states become degenerate and form together the ground-state manifold.
\label{fig:twopara}
} 
\end{figure*}

The major $q=0$ magnetic orders arise as a result of the degeneracy lifting by various anisotropic interactions and 
it is shown above that $D>0$, $K<0$, $\Gamma <0$ would favor the $\mathbb{Z}_2$ windmill, $\mathbb{Z}_6^{2p}$, $\mathbb{Z}_6^{1p}$ states,
respectively, when they are separately present in addition to the Heisenberg interaction.
Now we consider the cases where two different anisotropies exist in addition to the Heisenberg interaction
and investigate the interplay between two competing degeneracy breaking perturbations.

We present the phase diagrams of the $J$-$K$-$D$, $J$-$K$-$\Gamma$, $J$-$D$-$\Gamma$ models in Fig. \ref{fig:twopara}.
Here we again focus on the $\mathbb{Z}_2$ windmill (orange), $\mathbb{Z}_6^{2p}$ (pink), $\mathbb{Z}_6^{1p}$ (light blue) states.
Notice that no other $q=0$ state arises in the phase diagram.
First, the $\mathbb{Z}_2$ windmill states are generally favored when $K<0$, $D>0$, ${\Gamma}>0$.
On the other hand, the $\mathbb{Z}_6^{2p}$ states are stabilized when $K <0$, $D<0$, $\Gamma<0$, especially with
comparable magnitudes of $D$ and $\Gamma$.
The $\mathbb{Z}_6^{1p}$ states are found to appear when the symmetric \& anisotropic interaction $\Gamma<0$ is 
dominant over other anisotropies.

As discussed earlier, the $\mathbb{Z}_2$ windmill and $\mathbb{Z}_6^{2p}$ states are the degenerate ground-states when only the 
ferromagnetic Kitaev interaction is present in addition to the Heisenberg interaction [denoted with the yellow line in Figs. \ref{fig:twopara} (a) and (b)].
It is interesting to note that this $\mathbb{Z}_8$ degeneracy is lifted when the $J$-$K$ model with $K<0$ is additionally 
perturbed by the $D$ or $\Gamma$ interaction: the positive $D$ and $\Gamma$ favor the $\mathbb{Z}_2$ states 
while the opposite sign choices select the $\mathbb{Z}_6^{2p}$ states [see the orange and pink regions in Figs. \ref{fig:twopara} (a) and (b)].
Such competition between the $\mathbb{Z}_2$ and $\mathbb{Z}_6^{2p}$ states was discussed in a previous study on the classical $J$-$K$-$D$ model.\cite{Shindou2016}

Apart from the noncoplanar $q=0$ states, we find other magnetic phases such as 
the ferrimagnetic (green), N\'eel (gray), and incommensurate (white) phases.
Among these, the incommensurate state occupies a large region in the phase diagram,
reflecting the magnetic frustration arising from the competing anisotropic interactions.

\section{Full phase diagram}
We now discuss the full phase diagrams of the generic $J$-$K$-$\Gamma$-$D$ model shown in Fig. \ref{fig:JKDG}.
Here, the coupling constants are parametrized as follows.
\begin{subequations}
\begin{eqnarray}
J&=&1,
\\
K&=&0.3~\textup{sin}\theta~\textup{cos}\phi,
\\
\Gamma&=&0.3~\textup{sin}\theta~\textup{sin}\phi,
\\
D&=&0.3~\textup{cos}\theta,
\end{eqnarray}
\label{eq:parametrization}
\end{subequations}
where $0 \leq \theta < \pi$ and $0 \leq \phi < 2\pi$.
The two diagrams in the figure correspond to the two different signs of the Dzyaloshinskii-Moriya interaction: (a) $D>0$ and (b) $D<0$.
In each case, the center and the circumference of the disk diagram represent the $J$-$D$ ($\theta=0$ or $\pi$) and 
$J$-$K$-$\Gamma$ ($\theta=\pi/2$) models, respectively.
These two limits are interpolated by moving along the radial direction (parametrized by $\theta$).
The circumferential direction is represented by the other angular variable $\phi$.

As shown in the phase diagram, the noncoplanar $q=0$ states appear as dominant commensurate phases even when all three anisotropies come into play together.
Notably, for the positive DM coupling ($D>0$), the windmill state prevails in the vast region connecting the $J$-$K$-$\Gamma$ and the $J$-$D$ models
and pushes away the incommensurate phase from the center.
When the DM coupling is negative ($D>0$), the windmill state, however, loses its dominance over the incommensurate phase.
The latter extends from the circumference ($J$-$K$-$\Gamma$ model) to the center ($J$-$D$ model) when $D>0$.
Other phases such as the ferrimagnetic and N\'eel states show up as a point and in a small region of the phase diagrams.

\section{Discussion}

In this work, we constructed a generic local-moment model for the hyperkagome iridate Na$_4$Ir$_3$O$_8$, which includes
various frustrating anisotropic interactions ($K$, $\Gamma$, $D$) between the $j_{\textup{eff}}=1/2$ moments in addition to the dominant Heisenberg interaction ($J$).
Using the Luttinger-Tisza analysis and simulated annealing, we mapped out the classical phase diagram.
It is found that there exist three dominant $q=0$ noncoplanar magnetic orders as well as an incommensurate order.
The $q=0$ orders ($\mathbb{Z}_2$, $\mathbb{Z}_6^{2p}$, $\mathbb{Z}_6^{1p}$) are characterized by discrete degeneracies and
the degenerate classical ground-states are related to each other via the global $C_3$ rotation and/or the time-reversal symmetry.

We compare our work with a recent study on the $J$-$K$-$D$ model in Ref. [\onlinecite{Shindou2016}].
When $K$ and $D$ are both negative, the model has the $\mathbb{Z}_6^{2p}$ states as shown in Fig. \ref{fig:twopara} (a).
In Ref. [\onlinecite{Shindou2016}], it was claimed that the $\mathbb{Z}_6^{2p}$ states are selected by thermal order-by-disorder effect at low temperatures.
However, we find that the $\mathbb{Z}_6^{2p}$ states are readily stabilized at the zero temperature by the interplay of the anisotropic interactions.
Our results imply that the $\mathbb{Z}_6^{2p}$ states remain stable above zero temperature, and their stability is driven by energetics, 
namely the anisotropic interactions.

\begin{figure}[!htb]
\begin{center}
\includegraphics[width=7cm]{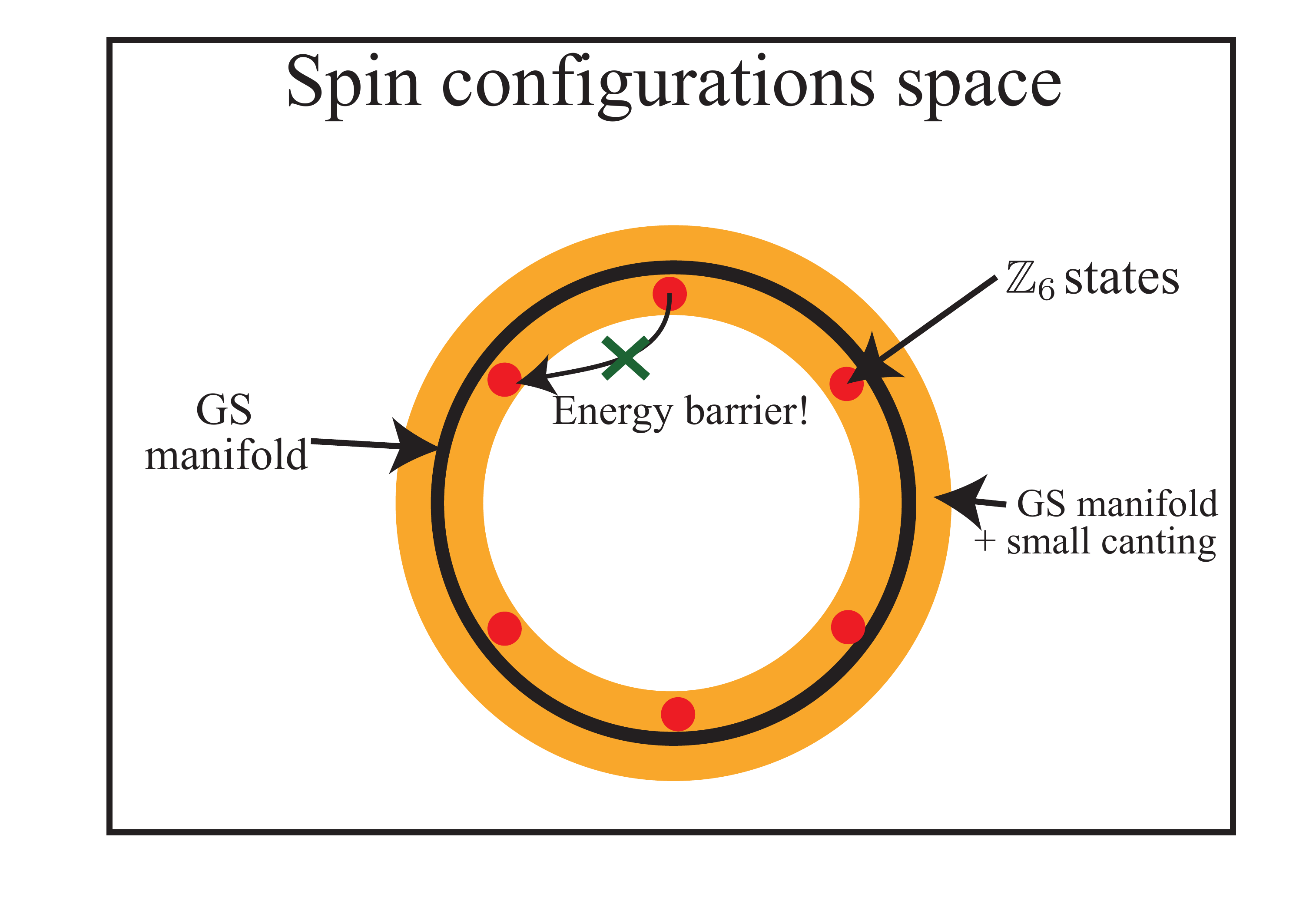}
\caption{
(Color online) Schematic picture for the degenerate ground-state manifold of the pure Heisenberg model (black circle),
ground-state manifold plus small canting components (orange shade),
and $\mathbb{Z}_6$ states (red dots)
in the spin configuration space.
 } \label{fig:dynamics}
\end{center} 
\end{figure}

Now we discuss possible implications of the $q=0$ orders for the recent $\mu$SR and NMR experiments.\cite{Dally2014,Shockley2015}
In these experiments, spin freezing behaviors or slow spin fluctuations have been discovered below $T_f = 6$-7 K in polycrystalline samples.
Assuming that the short-range magnetic orders below 6-7 K are determined by various anisotropic interactions, the $q=0$ magnetic orders 
$\mathbb{Z}_6^{2p}$ and $\mathbb{Z}_6^{1p}$, if they are taken as the dominant short-range magnetic correlations, may offer an explanation 
for the spin freezing behaviors. 
In the high-temperature regime
($K,D,\Gamma<T<J$),
the spin dynamics in the presence of
thermal and quantum fluctuations is constrained 
to occur near the degenerate ground-state manifold of the Heisenberg limit
(denoted by the orange shade in Fig. \ref{fig:dynamics}). 
Upon lowering the temperature,
the effects of anisotropic interactions become important 
and the system sees discrete shallow energy minima in the ground-state manifold representing six spin configurations in $\mathbb{Z}_6^{2p}$/ $\mathbb{Z}_6^{1p}$
(denoted by the red dots in Fig. \ref{fig:dynamics}).
We can then expect that a short-range correlation starts to form, 
which means that the system may be locally trapped in one of the discrete energy minima.
Since fluctuations to the other energy minima through the ground-state manifold of the Heisenberg limit are highly suppressed, the system exhibits spin freezing behaviors (or slow spin dynamics).

We support the above idea by showing that there is a large kinetic barrier between any pair of the six degenerate $\mathbb{Z}_6^{2p}$/ $\mathbb{Z}_6^{1p}$ states.
Starting from one state, we rotate spins one-by-one to reach another member of the six degenerate spin states.
We find that the energy barrier between two degenerate states scales linearly with the system size (for the energy barrier calculation; see Appendix \ref{app:energy_barrier}).
This implies that due to the large kinetic barrier six degenerate spin states are essentially disconnected and it is hard to move from one spin state to another.
Hence different kinds of degenerate spin states with
short-range order may persist for a long period of time in different regions of the system. 
This can lead to spin freezing behaviors or slow spin fluctuations.

One can imagine that quantum fluctuations may overcome the kinetic barrier at low temperatures and restore the locally-broken
(by the $q=0$ short-range order) $C_3$ symmetry. In this case, the quantum ground-state may form a quantum spin liquid with
global $C_3$ symmetry. Finding such a quantum ground-state and making connections to the classical limit would be an excellent 
topic of future research.

\acknowledgements
We are grateful to P. Mendels for detailed explanations of his experimental data and helpful discussions. 
This work was supported by the NSERC of Canada, the Canadian Institute for Advanced Research, and the Center for Quantum Materials
at the University of Toronto. This research was also supported in part by Perimeter Institute for Theoretical Physics. 
Research at Perimeter Institute is supported by the Government of Canada through Industry Canada and 
by the Province of Ontario through the ministry of Research and Innovation.
T. M. is supported by Advanced Leading Graduate Course for Photon Science (ALPS).

\appendix
\renewcommand{\thetable}{\Alph{section}.\arabic{figure}}
\renewcommand{\thefigure}{\Alph{section}.\arabic{table}}

\section{Derivation of the spin exchange interactions\label{sec:modelderivation}}

In this appendix, we provide the derivation of the $j_{\textup{eff}}=1/2$ spin model [Eq. (\ref{eq:Hamiltonian})] for the hyperkagome iridate Na$_4$Ir$_3$O$_8$.
First, we briefly review the strong-coupling expansion for the Kanamori type multiorbital interactions.\cite{Kanamori1963,Rau2014,Sugano1970,Rauthesis}
Then, we construct our model for Na$_4$Ir$_3$O$_8$.

\subsection{Strong-coupling expansion}

\begin{table}
\begin{ruledtabular}
\begin{tabular}{c|cc|c}
$N_i$ & $S_i$ & $L_i$ & Energy
\\
\hline
0 & 0 & 0 & $(U-3J_H)/2$
\\
\hline
1 & 1/2 & 1 & $-5J_H/2$
\\
\hline
2 & 0 & 0 & $(U-3J_H)/2$
\\
2 & 1 & 1 & $(U-13J_H)/2$
\\
2 & 0 & 2 & $(U-9J_H)/2$
\end{tabular}
\end{ruledtabular}
\caption{Eigenstates of the Kanamori Hamiltonian $H_{\mathrm{int}}$.
The eigenstates are characterized with the the total hole number ($N_i$), spin ($S_i$), and angular momentum ($L_i$).
}
\label{tab:Kanamori_eigen}
\end{table}

For a simple setting of the strong-coupling expansion, we consider a two-site system described by the following Hamiltonian.
\begin{equation}
H_{12} = H_{\mathrm{int}} + H_{\mathrm{soc}} + H_{\mathrm{hop}}.
\end{equation}
We assume that each site has five electrons (or one hole) in the $t_{2g}$ manifold as in the Ir$^{4+}$ ion.
The above Hamiltonian consists of the atomic multiorbital interactions ($H_{\mathrm{int}}$), spin-orbit coupling ($H_{\mathrm{soc}}$), and electron hoppings ($H_{\mathrm{hop}}$) between the two sites.
In the strong-coupling expansion, we assume that $H_{\mathrm{int}} \gg H_{\mathrm{soc}} \gg H_{\mathrm{hop}}$.

First, we consider the system in the atomic limit described by only the interaction term:
\begin{equation}
H_{\mathrm{int}} = \sum_{i=1}^2 \frac{U-3J_H}{2}(N_i-1)^2 - 2J_H \bm{S}_i^2 - \frac{J_H}{2} \bm{L}_i^2.
\end{equation}
Here, we employed the well known Kanamori Hamiltonian for the multiorbital interactions.
The Hamiltonian is parametrized with the intraorbital Coulomb interaction ($U$) and the Hund's coupling ($J_H$).
It is readily diagonalized in terms of the total hole number ($N_i$) in the $t_{2g}$ manifold, and total spin ($\bm{S}_i$) and total orbital ($\bm{L}_i$) angular momenta at each site ($i=1,2$).
Hence, the eigenstates at each site can be represented by the three quantum numbers: $|N_i,S_i,L_i\rangle$. 
Notice that we are using the hole basis here instead of the electron basis.
We summarize the eigenstates with $N_i=0,1,2$ in Table \ref{tab:Kanamori_eigen}.
The atomic ground-state manifold is characterized with the quantum numbers: $N_i=1,~S_i=1/2,~L_i=1$. 

Next, we turn on the atomic spin-orbit coupling:
\begin{equation}
H_{\mathrm{soc}}=\sum_{i=1}^2 -\lambda \bm{L}_i \cdot \bm{S}_i.
\end{equation}
Note that the coupling constant ($-\lambda$) has a minus sign when written in the hole basis.
We incorporate effect of the spin-orbit coupling by projecting the atomic ground-state manifold $|N_i=1,S_i=1/2,L_i=1\rangle$ into the $j_{\textup{eff}}=1/2$ Kramers doublet (where $\bm{j}_{\textup{eff}}=\bm{L}+\bm{S}$).

We introduce electron/hole hoppings between the two sites.
We assume the most generic hopping Hamiltonian as follows.
\begin{equation}
H_{\mathrm{hop}} = \sum_{\sigma}\hat{d}^{\dagger}_{1,\sigma} t_{12} \hat{d}_{2,\sigma}
+ \textup{H.c.}
\end{equation}
Here, $\hat{d}_{i,\sigma} = (d_{i,yz,\sigma},d_{i,xz,\sigma},d_{i,xy,\sigma})^\mathrm{T}$ are the hole annihilation operators at the site $i$.
The subscripts $yz,~xz,~xy$ represent the single-hole ($N_i=1$) states in the $t_{2g}$ basis, and $\sigma(=\uparrow,\downarrow)$ means the spin state of the hole.
The hopping amplitude matrix $t_{12}$ is parametrized with nine independent real parameters:
\begin{equation}
t_{12} =
\left(
\begin{array}{ccc}
s+q_{xx} & q_{xy}  +v_z&  q_{xz}-v_y \\
q_{xy}-v_z & s+q_{yy} & q_{yz}+v_x \\
q_{xz}+v_y  & q_{yz} -v_x& s+q_{zz} \\
\end{array}
\right).
\label{eq:hopping_mat}
\end{equation}
The hopping matrix is basically decomposed into the trace ($s$), antisymmetric vector ($\bm{v}$), and traceless symmetric matrix ($\bm{q}$).

Now the effective exchange interactions are derived with the strong-coupling expansion.
As mentioned earlier, we assume that $U,J_H \gg \lambda \gg t$.
By reflecting the hopping effects on the $j_{\textup{eff}}=1/2$ doublets via the second order perturbation theory, we obtain the effective exchange interactions:
\begin{equation}
\mathcal{H}_{12} = 
\bm{S}_1^{\mathrm{T}}
\left(
\begin{array}{ccc}
 \tilde{J} + \tilde{\Gamma}_{xx}  & \tilde{\Gamma}_{xy}  +\tilde{D}_z & \tilde{\Gamma}_{xz} -\tilde{D}_y \\
 \tilde{\Gamma}_{xy} - \tilde{D}_z & \tilde{J} +\tilde{\Gamma}_{yy} & \tilde{\Gamma}_{yz}+ \tilde{D}_x \\
\tilde{ \Gamma}_{xz} +\tilde{D}_y & \tilde{\Gamma}_{yz} -\tilde{D}_x & \tilde{J} + \tilde{\Gamma}_{zz}  \\
 \end{array}
 \right)
\bm{S}_2 .
\label{eq:Hamiltonian_2}
\end{equation}
It must be noted that here the operators $\bm{S}_{1,2}$ are the $j_{\textup{eff}}=1/2$ pseudospin operators at the sites, 1 and 2.
The coupling constants are given by the following expressions.
\begin{widetext}
\begin{subequations}
\label{eq:JDG}
\begin{eqnarray}
\tilde{J} 
&=& 
\frac{4}{27} \left[ \frac{18s^2 - \bm{v}^2}{U-3J_H} 
- \frac{\frac{5}{3}\bm{v}^2}{U-J_H} + \frac{9s^2 -\frac{4}{3}\bm{v}^2}{U+2J_H}
- \frac{3 J_H \mathrm{tr}(\bm{q}^2)}{(U-3J_H)(U-J_H)} \right],
\\
\tilde{\bm{D}}
&=& 
-\frac{16}{9} \left[ \left(\frac{2}{U-3J_H} 
+ \frac{1}{U+2J_H}\right) s\bm{v}
+ \frac{J_H \bm{q} \bm{v} }{(U-3J_H)(U-J_H)}\right] ,
\\
\tilde{\bm{\Gamma} }
&=& 
\frac{4}{27} 
\left[ \left(\frac{15}{U-3J_H} + \frac{1}{U -J_H} 
+\frac{8}{U+2J_H}\right) \left(\bm{v}\bm{v}^{\mathrm{T}}-\frac{\bm{v}^2}{3}\right)
+ \frac{18J_H (\bm{q}^2 -\frac{1}{3}\mathrm{tr}(\bm{q}^2) )}{(U-3J_H)(U-J_H)} \right].
\end{eqnarray} 
\end{subequations}
\end{widetext}
One can easily check that the $\tilde{\bm{\Gamma}}$ matrix is traceless: $\tilde{\Gamma}_{xx}+\tilde{\Gamma}_{yy}+\tilde{\Gamma}_{zz}=0$.

\subsection{Model for the hyperkagome iridate}
\begin{figure}
\centering
\includegraphics[width=7cm]{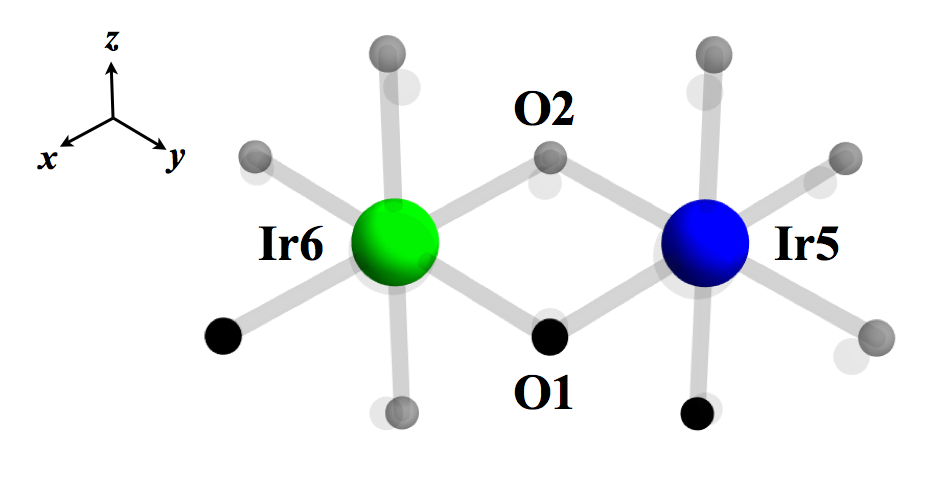}
\caption{
(Color online) The idealized crystal structure.
The figure depicts local environment of the Ir 5 and 6 sites (Fig. \ref{fig:lattice}) in the idealized crystal structure.
Here each Ir-O bond is parallel to one of the global $x,y,z$ axes and the bond length is uniform across all the bonds. The O$_1$ (black) and O$_2$ (gray) sites are distinguished by different local environment: O$_1$ (O$_2$) sites are shared by neighboring three (two) Ir ions.
For comparison, the actual crystal structure of Na$_4$Ir$_3$O$_8$\cite{Okamoto2007} is drawn together with faint gray balls.
} 
\label{fig:IrO6}
\end{figure}

We construct the $j_{\textup{eff}}=1/2$ spin model for the hyperkagome iridate Na$_4$Ir$_3$O$_8$.
In the actual crystal structure of Na$_4$Ir$_3$O$_8$,\cite{Okamoto2007} most of the anisotropic couplings in $\mathcal{H}_{12}$ [Eq. (\ref{eq:Hamiltonian_2})] are expected to be nonzero due to lattice distortions from an ideal structure.
In this work, instead of pursuing the actual crystal structure, we idealize the structure in such a way that each Ir-O bond is parallel to one of the global $x,y,z$ axes and the bond length is uniform across all the bonds (Fig. \ref{fig:IrO6}).
With the idealized crystal structure, we derive a relatively simple, but still generic spin Hamiltonian for Na$_4$Ir$_3$O$_8$.
We show the derivation by taking the Ir-Ir bond (6,5) in Fig. \ref{fig:lattice} as an example.
First, we express the hopping matrix at the bond by using the Slater-Koster parametrization:\cite{Slater1954}
\begin{align}
& T_{65}  \notag \\ 
=& 
\left(
\begin{array}{ccc}
 \frac{V_{dd\delta}+ V_{dd\pi}}{2}  &\frac{V_{dd\delta}-V_{dd\pi}}{2}  -\frac{ V_{pd \pi 1}^2}{\Delta_1}& 0 \\
\frac{V_{dd\delta}-V_{dd \pi}}{2} -\frac{V_{pd \pi 2}^2}{\Delta_2} & \frac{V_{dd\delta}+V_{dd\pi}}{2}& 0 \\
 0 & 0& \frac{V_{dd\delta}}{4}+\frac{3 V_{dd\sigma}}{4}\\
\end{array}
\right).
\label{eq:HM_simple}
\end{align} 
Here, the parameters $V_{dd\sigma}$, $V_{dd\pi}$, $V_{dd\delta}$ represent direct hoppings between the neighboring Ir sites.
The parameter $V_{pd \pi i}~(i=1,2)$ means the hoppings between the $2p$ orbitals at the O$_i$ site 
and $t_{2g}$ orbitals at an Ir site, and $\Delta_i$ implies the energy difference between the former and latter orbitals.
Hence, the amplitude $-\frac{ V_{pd \pi i}^2}{\Delta_i}$ indicates the indirect hopping via the intermediate oxygen site O$_i$.
Notice that there are two distinct oxygen sites, O$_1$ and O$_2$, in the compound (see Fig. \ref{fig:IrO6}).
The O$_1$ (O$_2$) sites are shared by neighboring three (two) Ir ions.
We assume that $V_{pd \pi 1} \neq V_{pd \pi 2}$ and $\Delta_1 \neq \Delta_2$ for the inequivalent O$_1$ and O$_2$ sites.
The above hopping matrix can be decomposed into the form of Eq. (\ref{eq:hopping_mat}) with the following nonzero parameters: 
\begin{subequations}
\label{eq:svq}
\begin{eqnarray}
s &=& \frac{3V_{dd\sigma} + 4V_{dd\pi} +  V_{dd\delta} }{12},
\\
v_z &=& \frac{V_{pd\pi 2}^2}{2\Delta_2}-\frac{V_{pd\pi 1}^2}{2\Delta_1}, 
\\ 
q_{xx} &=& q_{yy} = -2 q_{zz}
\nonumber\\ 
&=&
\frac{3V_{dd\sigma} + 2V_{dd\pi} -V_{dd\delta}}{12},
\\
q_{xy} 
&=&
\frac{V_{dd\delta}-V_{dd\pi} }{2} - \frac{V_{pd\pi 1}^2}{2\Delta_1}-\frac{V_{pd\pi 2}^2}{2\Delta_2}.
\end{eqnarray}
\end{subequations}

The resulting exchange interactions caused by the above hopping amplitudes take the following form.
\begin{equation}
\mathcal{H}_{65} = 
\bm{S}_6^{\mathrm{T}}
\left(
\begin{array}{ccc}
 \tilde{J} -\frac{\tilde{\Gamma}_{zz}}{2}  & \tilde{\Gamma}_{xy}  +\tilde{D}_z &0 \\
 \tilde{\Gamma}_{xy} - \tilde{D}_z & \tilde{J}  -\frac{\tilde{\Gamma}_{zz}}{2}& 0 \\
0 & 0 & \tilde{J} + \tilde{\Gamma}_{zz}  \\
 \end{array}
 \right)
\bm{S}_5.
\label{hamiltonian_3}
\end{equation}
Microscopic expression for the coupling constants $\tilde{J}$, $\tilde{D}_z$, $\tilde{\Gamma}_{zz}$ and $\tilde{\Gamma}_{xy}$ can be obtained by plugging Eq. (\ref{eq:svq}) into (\ref{eq:JDG}).
Lastly, we simplify the above bond Hamiltonian into the final form:
\begin{align}
\mathcal{H}_{65} = 
\bm{S}_{6}^{\mathrm{T}}
\left(
\begin{array}{ccc}
J & -\Gamma + D & 0 \\
-\Gamma -D & J & 0 \\
0 & 0 & J+K
\end{array}
\right)
\bm{S}_{5}.
\end{align}
Here, the compling constants for the Heisenberg ($J$), Kitaev ($K$), Dzyaloshinskii-Moriya ($D$), and anisotropic \& symmetric ($\Gamma$) interactions are defined as follows.
\begin{subequations}
\begin{eqnarray}
J &=& \tilde{J}-\tilde{\Gamma}_{zz}/2,
\\
K &=& 3\tilde{\Gamma}_{zz}/2,
\\
D &=& \tilde{D}_z,
\\
\Gamma &=& -\tilde{\Gamma}_{xy}.
\end{eqnarray}
\end{subequations}
The exchange interactions at other bonds are generated by applying the $C_3$ and $C_2$ symmetry operations (Sec. \ref{sec:lattice}) to the bond Hamiltonian $\mathcal{H}_{65}$.
Then, we obtain the model Hamiltonian $\mathcal{H}$ [Eq. (\ref{eq:Hamiltonian})].

\section{Luttinger-Tisza analysis for the $J$-$K$ model\label{sec:LTA-JK}}

The Luttinger-Tisza analysis (LTA) for the $J$-$K$ model is discussed in details here.
First, the LTA is briefly reviewed.\cite{Luttinger1946,Luttinger1951}
In the LTA, we relax the hard spin constraint $|\bm{S}_{i}| = 1$ and implement it on average: $| \bm{S}_{1} + \bm{S}_{2} + \cdots + \bm{S}_{N} | = N$ ($N$ is the number of the spin moments).
The resulting quadratic Hamiltonian matrix is solved in the momentum space.
\begin{equation}
\mathcal{H}= \sum_{\bm{q}} \bm{S}^{\mathrm{T}} (-\bm{q}) \mathcal{J}(\bm{q}) \bm{S}(\bm{q}),
\end{equation}
In this expression, the $3N_s\times3N_s$ matrix $\mathcal{J}(\bm{q})$ is the block Hamiltonian matrix in the momentum $\bm{q}$ sector ($N_s$ is the number of sublattices in a unit cell, and $N_s=12$ in our hyperkagome lattice model).
The $3N_s$-component column vector $\bm{S}(\bm{q})$ represents 
a Fourier component of real-space spins $(\bm{S}_1,\bm{S}_2,\cdots,\bm{S}_N)$.
After finding the lowest-energy state of $\mathcal{H}$,
we check whether the state satisfies the hard spin constraint.
If it does, the lowest-energy state is the {\it exact} ground-state of the Hamiltonian.
When the hard spin constraint is not satisfied, the LTA provides a lower bound for the ground-state energy.

\begin{figure*}[!htb]
\begin{center}
\includegraphics[width=17cm]{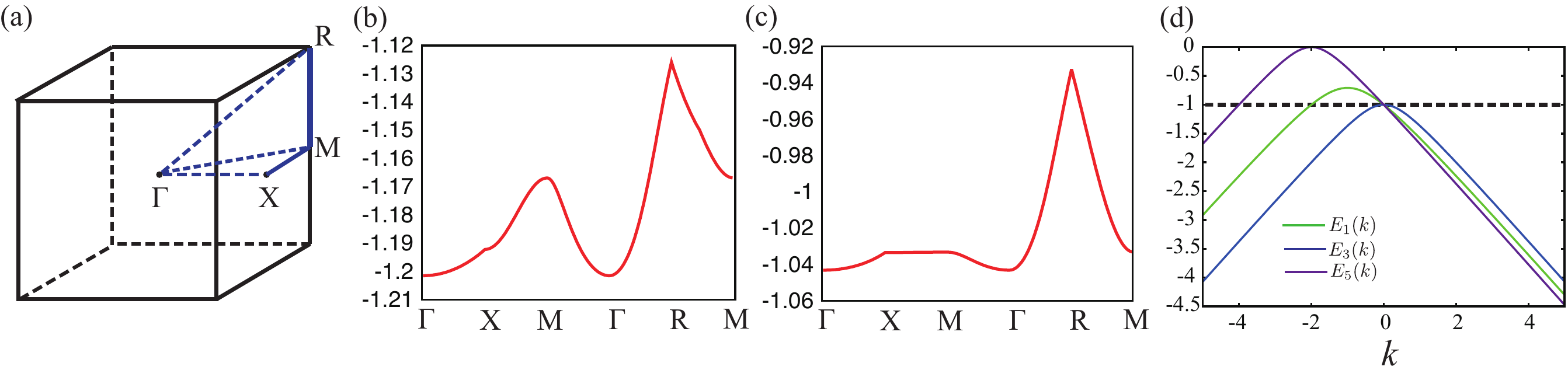}
\caption{(a) The first Brillouin zone of the cubic lattice and the high-symmetry points, 
the LT band structures for $J$-$K$ model with (b) $K=0.3J$
and (c) $K=-0.3J$,
and (d) $k$ dependence of $E_1(k), E_3(k)$ and $E_5(k)$.
 } \label{fig:LTA-JK}
\end{center} 
\end{figure*}

Now we apply the LTA to the $J$-$K$ model $\mathcal{H}_{JK}$.
One can easily find that the lowest-energy state occur at $\bm{q}=0$ by diagonalizing the Hamiltonian matrix $\mathcal{J}_{JK}(\bm{q})$ (see Fig. \ref{fig:LTA-JK}).
Hence, we focus on the $\bm{q}=0$ sector of $\mathcal{H}_{JK}$ and analyze the spin structure of the lowest-energy mode.
The $\bm{q}=0$ Hamiltonian can be block-diagonalized in the following way:
\begin{equation}
\mathcal{H}_{JK}(\bm{q}=0) = 
\left(
\begin{array}{ccc}
\bm{S}_x^{T} & \bm{S}_y^{T} & \bm{S}_z^{T}
\end{array}
\right)
\left(
\begin{array}{ccc}
\mathcal{J}_{x} &0  & 0\\
0 & \mathcal{J}_y & 0\\
0 & 0 & \mathcal{J}_z
\end{array}
\right)
\left(
\begin{array}{c}
\bm{S}_x\\ 
\bm{S}_y\\
\bm{S}_z
\end{array}
\right).
\end{equation}
Here, we take the basis for the spin vector as follows.
\begin{eqnarray}
\bm{S}_{x}^T
&=&
(S_6^x,S_4^x,S_2^x,S_5^x,S_1^x,S_3^x,S_{8}^x,S_{10}^x,S_{12}^x,S_{9}^x,S_{7}^x,S_{11}^x),
\nonumber\\
\bm{S}_{y}^T
&=&
(S_{12}^y,S_5^y,S_4^y,S_1^y,S_2^y,S_{9}^y,S_6^y,S_{11}^y,S_{10}^y,S_3^y,S_{8}^y,S_{7}^y),
\nonumber\\
\bm{S}_{z}^T
&=&
(S_{7}^z,S_6^z,S_5^z,S_2^z,S_3^z,S_{10}^z,S_1^z,S_{12}^z,S_{11}^z,S_{8}^z,S_{9}^z,S_4^z).
\nonumber\\
\end{eqnarray}
The three 12$\times$12 matrices $\mathcal{J}_{x,y,z}$ are given by 
\begin{eqnarray}
\mathcal{J}_x
&=& 
\left(
\begin{array}{cccc}
D & C & 0 & B\\
C^\mathrm{T} & D & A & 0\\
 0 & A & D & C^\mathrm{T} \\
B & 0 & C & D 
\end{array}
\right),
~~~
\mathcal{J}_y
=
\left(
\begin{array}{cccc}
D & C & B & 0 \\
C^\mathrm{T} & D & 0 & C  \\
B & 0  & D & A \\
0 & C^\mathrm{T}  & A & D 
\end{array}
\right),
\nonumber\\
\mathcal{J}_z
&=& 
\left(
\begin{array}{cccc}
D & 0 & A & C^\mathrm{T} \\
0 & D & C & B\\
A & C^\mathrm{T} & D & 0 \\
C & B  &0& D 
\end{array}
\right)
\end{eqnarray}
with the submatrices 
\begin{eqnarray}
A 
&=& 
\left(
\begin{array}{ccc}
0 & 0 & \frac{1}{2} \\
0 & 0 & \frac{1}{2}\\
\frac{1}{2} & \frac{1}{2} & 0 
\end{array}
\right),
~~~
B 
= 
\left(
\begin{array}{ccc}
0 & \frac{1}{2} & \frac{1}{2} \\
\frac{1}{2} & 0 & 0 \\
\frac{1}{2} & 0 & 0 
\end{array}
\right),
\nonumber\\
C 
&=& 
\left(
\begin{array}{ccc}
\frac{1}{2} & 0 & 0 \\
\frac{1}{2} & 0 & 0 \\
0 & \frac{1}{2} & \frac{1}{2} 
\end{array}
\right),
~~~
D 
=
\left(
\begin{array}{ccc}
0 & \frac{1+k}{2} & 0 \\
\frac{1+k}{2} & 0  &\frac{1+k}{2}\\
0 & \frac{1+k}{2} & 0 
\end{array}
\right).
\end{eqnarray}
Here, we use the reduced coupling constant $k~(=K/J)$.
Notice that $\mathcal{J}_{x,y,z}$ are equivalent matrices connected by unitary transformations.
The twelve energy eigenvalues shared by the three matrices are obtained as follows.
\begin{itemize}
\item $\pm$1
\item 0 (2-fold)
\item $E_1(k) = -\sqrt{\frac{2+k(2+k)}{2}}$ (2-fold)
\item $E_2(k) =\sqrt{\frac{2+k(2+k)}{2}}$ (2-fold)
\item $E_3(k) =-\frac{1+\sqrt{1+2k^2}}{2}$ 
\item $E_4(k) =\frac{-1+\sqrt{1+2k^2}}{2}$ 
\item $E_5(k) =\frac{1-\sqrt{9+2k(4+k)}}{2}$ 
\item $E_6(k) =\frac{1+\sqrt{9+2k(4+k)}}{2}$ 
\end{itemize}
As shown in Fig. \ref{fig:LTA-JK} (d), a different lowest-energy state is selected by the Kitaev interaction depending on the sign of $k$. 

When the Kitaev interaction is antiferromagnetic ($k>0$), the ground-state has the energy $E_5(k)$ and the corresponding spin state is constructed in the following way.
First, we note that the three matrices $\mathcal{J}_{x,y,z}$ have the same eigenvector:
$\bm{S}_{x}^T=\bm{S}_{y}^T=\bm{S}_{z}^T \propto (\bm{u},\bm{u},\bm{u},\bm{u})$ with $\bm{u}=(1,-\frac{1+\sqrt{9 +8k +2k^2}}{2+k},1)$.
Then, we obtain the eight degenerate spin states by linearly combining the $\bm{S}_{x,y,z}$ as
\begin{equation}
\bm{S}
\propto
\left(
\begin{array}{c}
a ~ \bm{S}_x
\\
b ~ \bm{S}_y
\\
c ~ \bm{S}_z
\end{array}
\right)
\label{eq:linearcomb}
\end{equation}
with the sign factors $a,b,c~(=\pm)$.
One can check that the eight states satisfy the hard spin constraint. 
These states are the eightfold-degenerate ferrimagnetic states mentioned in the main text (Fig. \ref{fig:Z6_2p}).

In the case of the ferromagnetic Kitaev interaction ($k<0$), the matrices $\mathcal{J}_{x,y,z}$ have almost the same eigenvector except for the sign structure: $\bm{S}_{x}^T\propto (\bm{v},-\bm{v},\bm{v},-\bm{v})$, $\bm{S}_{y}^T\propto (-\bm{v},\bm{v},\bm{v},-\bm{v})$, $\bm{S}_{z}^T \propto (-\bm{v},-\bm{v},\bm{v},\bm{v})$ with $\bm{v} = (1,\frac{1-\sqrt{2k^2} +1}{|k|},1)$.
The ground-state manifold can be constructed by taking eight different combinations of $\bm{S}_{x,y,z}$, as we did in Eq. (\ref{eq:linearcomb}).
One can check the hard spin constraint for each of the eight states, and find that the ground-state manifold consists of the $\mathbb{Z}_2$ windmill states and $\mathbb{Z}_6^{2p}$ states (Fig. \ref{fig:Z6_2p}).
In this case, the ground-state energy is given by $E_3(k)$.

\section{Luttinger-Tisza analysis for the $J$-$\Gamma$ model\label{sec:LTA-JG}}
\begin{figure}[!htb]
\begin{center}
\includegraphics[width=8cm]{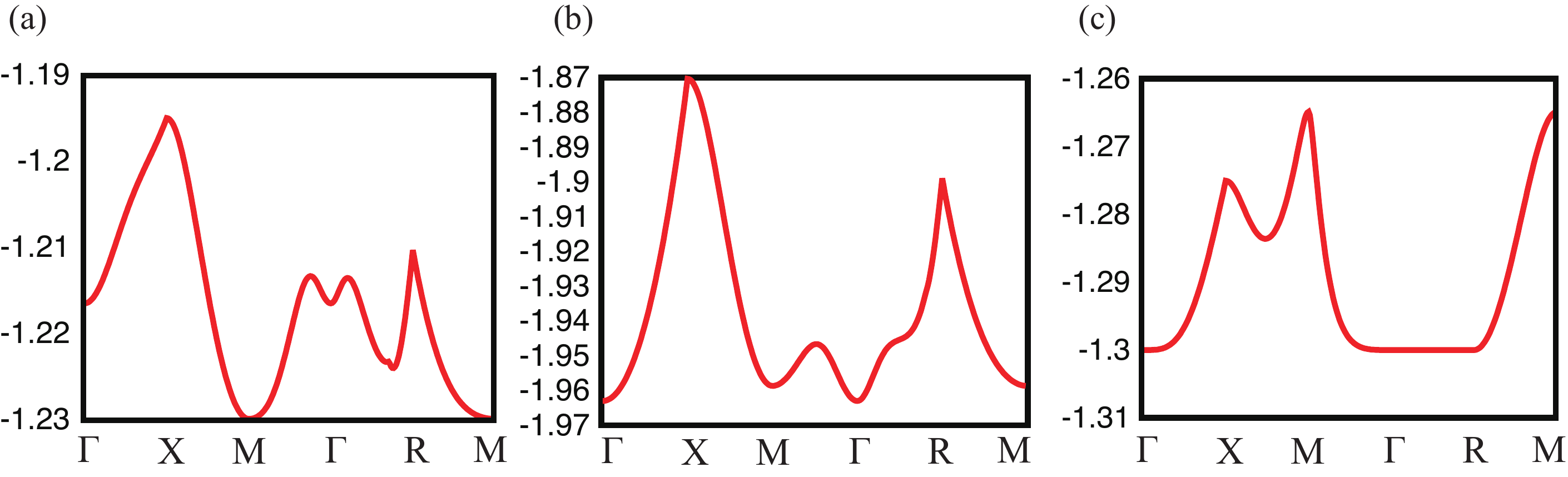}
\caption{Dispersion relation of the lowest band of J-$\Gamma$ model
with (a) $J=1,\Gamma = 0.2$, (b) $J=1, 
\Gamma = 0.8$, and $J=1, \Gamma=-0.3$.}
\label{fig:LTA-JG}
\end{center}
\end{figure}

In this appendix, we describe the LTA for the $J$-$\Gamma$ model $\mathcal{H}_{J\Gamma}$.
The lowest-energy state occurs at different positions in the Brillouin zone depending on the value of the coupling constant $\Gamma$ (see Fig. \ref{fig:LTA-JG}).
We focus on the $\Gamma<0$ case in which the LTA provides the exact ground-states of $\mathcal{H}_{J\Gamma}$.
In this case, the lowest-energy mode has a flat dispersion along the $\Gamma$R line in the Brillouin zone.
However, it turns out that none of the finite-$\bm{q}$ states satisfies the hard spin constraint (the absence of the finite-$\bm{q}$ states is also confirmed by our simulated annealing approaches).
In the following, we will examine the $\bm{q}=0$ states and construct the ground-state manifold for the $\Gamma<0$ case.

Interestingly, the $\bm{q}=0$ states are the lowest-energy states for each of $\mathcal{H}_J$ and $\mathcal{H}_{\Gamma}$.
The $\bm{q}=0$ states of $\mathcal{H}_{J\Gamma}$ can be obtained by investigating the ground-state manifold of $\mathcal{H}_{\Gamma}$ and then considering the Heisenberg interaction on the manifold.
Hence, we solve the $\Gamma$-only model $\mathcal{H}_{\Gamma}$ first.
We set $\Gamma=-1$ and block diagonalize the Hamiltonian matrix in the following fashion.
\begin{eqnarray}
&&
\mathcal{H}_\Gamma(\bm{q}=0)
\\
 &=&
\left(
\begin{array}{cccc}
\bm{S}_a^T &
\bm{S}_b^T &
\bm{S}_c^T &
\bm{S}_d^T
\end{array}
\right)
\left(
\begin{array}{cccc}
\mathcal{J}_{g} & 0 & 0 & 0 \\
0 & \mathcal{J}_{g} & 0 & 0 \\
0 & 0 & \mathcal{J}_{g} & 0 \\
0 & 0 & 0 & \mathcal{J}_{g}
\end{array}
\right)
\left(
\begin{array}{c}
\bm{S}_a \\
\bm{S}_b \\
\bm{S}_c \\
\bm{S}_d
\end{array}
\right),
\nonumber
\end{eqnarray}
where $\mathcal{J}_{g}$ is the 9$\times$9 matrix
\begin{equation}
\mathcal{J}_{g}
=
\left(
\begin{array}{ccc}
G & F  & F \\
F & G & F \\
F & F & G 
 \end{array}
 \right)
\end{equation}
with
\begin{equation}
G
=
\left(
\begin{array}{ccc}
0 & \frac{1}{2} & 0 \\ 
\frac{1}{2} & 0& \frac{1}{2}\\
0 & \frac{1}{2} & 0  
\end{array}
\right),
~~~~~
F 
=
\left(
\begin{array}{ccc}
\frac{1}{2} & 0 & 0\\ 
0 & 0 & 0 \\
0 & 0 & \frac{1}{2} 
\end{array}
\right).
\end{equation}
Here, as the basis for the spin vector we choose $\bm{S}^T=(\bm{S}_a^T,\bm{S}_b^T,\bm{S}_c^T,\bm{S}_d^T)$ with the following sequence of the spin components.
\begin{eqnarray}
\bm{S}_a^T &=& (S_6^x, S_5^y,S_1^z, S_{11}^z,S_{10}^x,S_3^y,S_7^y,S_9^z,S_2^{x}),
\nonumber\\
\bm{S}_b^T &=& (S_5^x, -S_{12}^z ,S_{10}^y , S_6^y ,S_7^x,- S_8^z,- S_4^z ,S_2^y, S_3^{x}),
\nonumber\\
\bm{S}_c^T &=& (S_5^z, -S_4^x ,S_9^y , S_1^y ,S_3^z,-S_8^x, -S_{12}^x ,S_{11}^y, S_{7}^{z}),
\nonumber\\
\bm{S}_d^T &=& (S_{9}^x, -S_{8}^y ,S_{10}^z , S_2^z  ,S_{1}^x,-S_{12}^y, -S_{4}^y ,S_{6}^z, S_{11}^{x}).
\end{eqnarray}
The lowest eigenvalue of $\mathcal{J}_g$ 
is $-1$ 
with the doubly degenerate eigenvectors, 
$\bm{w} = (-1,1,-1,0,0,0,1,-1,1)^T$
and 
$\bm{z} = (-1,1,-1,1,-1,1,0,0,0)^T$.
The ground-states satisfying the hard spin constraint can be constructed by combining the eigenvectors in the following ways.
\begin{equation}
\bm{S}
\propto
\left(
\begin{array}{c}
a~\bm{w}
\\
b~\bm{w}
\\
c~(\bm{w}-\bm{z})
\\
d~\bm{w}
\end{array}
\right),
~~~
\left(
\begin{array}{c}
a~(\bm{w}-\bm{z})
\\
b~\bm{z}
\\
c~\bm{z}
\\
d~(\bm{w}-\bm{z})
\end{array}
\right),
~~~
\left(
\begin{array}{c}
a~\bm{z}
\\
b~(\bm{w}-\bm{z})
\\
c~\bm{w}
\\
d~\bm{z}
\end{array}
\right).
\end{equation}
In this expression, the right hand side shows three different ways for the combinations with the sign factors $a,b,c,d~(=\pm)$.
Therefore, we find 3$\times$2$^4$=48 different states in the ground-state manifold of $\mathcal{H}_{\Gamma}$.

To obtain the ground-states of $\mathcal{H}_{J\Gamma}$, the Heisenberg interaction is considered on the ground-state manifold of $\mathcal{H}_{\Gamma}$.
By examining the Heisenberg interaction energy for each state, one can find that only six states in the manifold have the minimum energy, $-J/2$ per bond.
The six states represented by
\begin{equation}
\bm{S}
\propto
\pm
\left(
\begin{array}{c}
\bm{w}
\\
-\bm{w}
\\
(\bm{w}-\bm{z})
\\
\bm{w}
\end{array}
\right),
~~~
\pm
\left(
\begin{array}{c}
(\bm{w}-\bm{z})
\\
-\bm{z}
\\
\bm{z}
\\
-(\bm{w}-\bm{z})
\end{array}
\right),
~~~
\pm
\left(
\begin{array}{c}
\bm{z}
\\
(\bm{w}-\bm{z})
\\
-\bm{w}
\\
\bm{z}
\end{array}
\right),
\end{equation}
are the $\mathbb{Z}_6^{1p}$ states (Fig. \ref{fig:Z6_1p}).
The above three pairs of vectors sequently correspond to $\mathbb{Z}_6^{1p}$ $x,~y,~z$ states, respectively.
It is interesting to note that the state vectors do not depend on the coupling constant $\Gamma$.

\section{Spin structure factors for the $q=0$ magnetic orders\label{sec:SSF}}
\begin{figure}[!htb]
\begin{center}
\includegraphics[width=7cm]{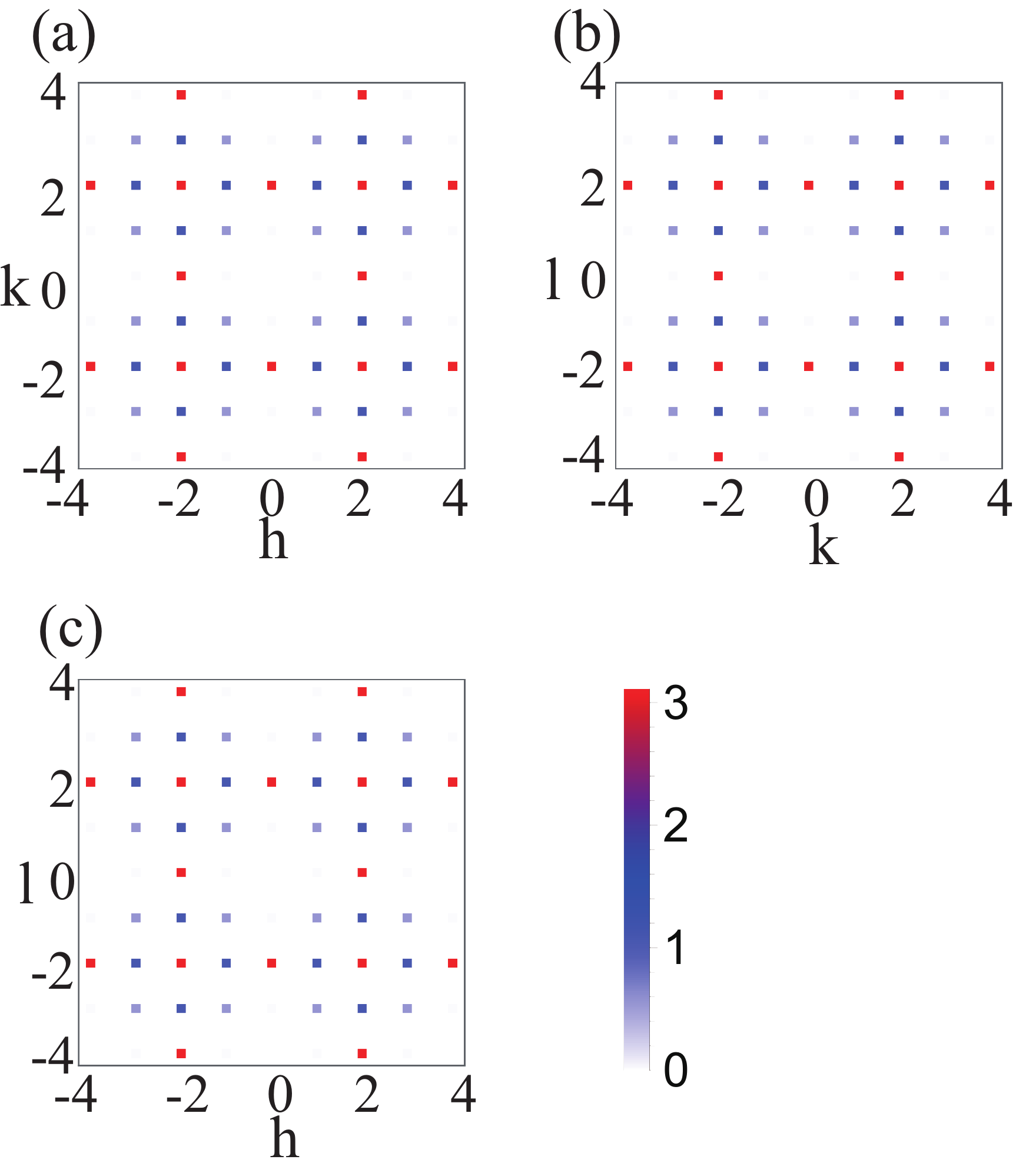}
\vspace{-10pt}
\caption{
(Color online) Spin structure factor of the $\mathbb{Z}_2$ wind-mill state
on the (a) $(hk0)$, (b) $(kl0)$ and (c) $(h0l)$ planes.} 
\label{fig:SF_Z2} 
\end{center}
\end{figure}

\begin{figure}[!htb]
\begin{center}
\includegraphics[width=7cm]{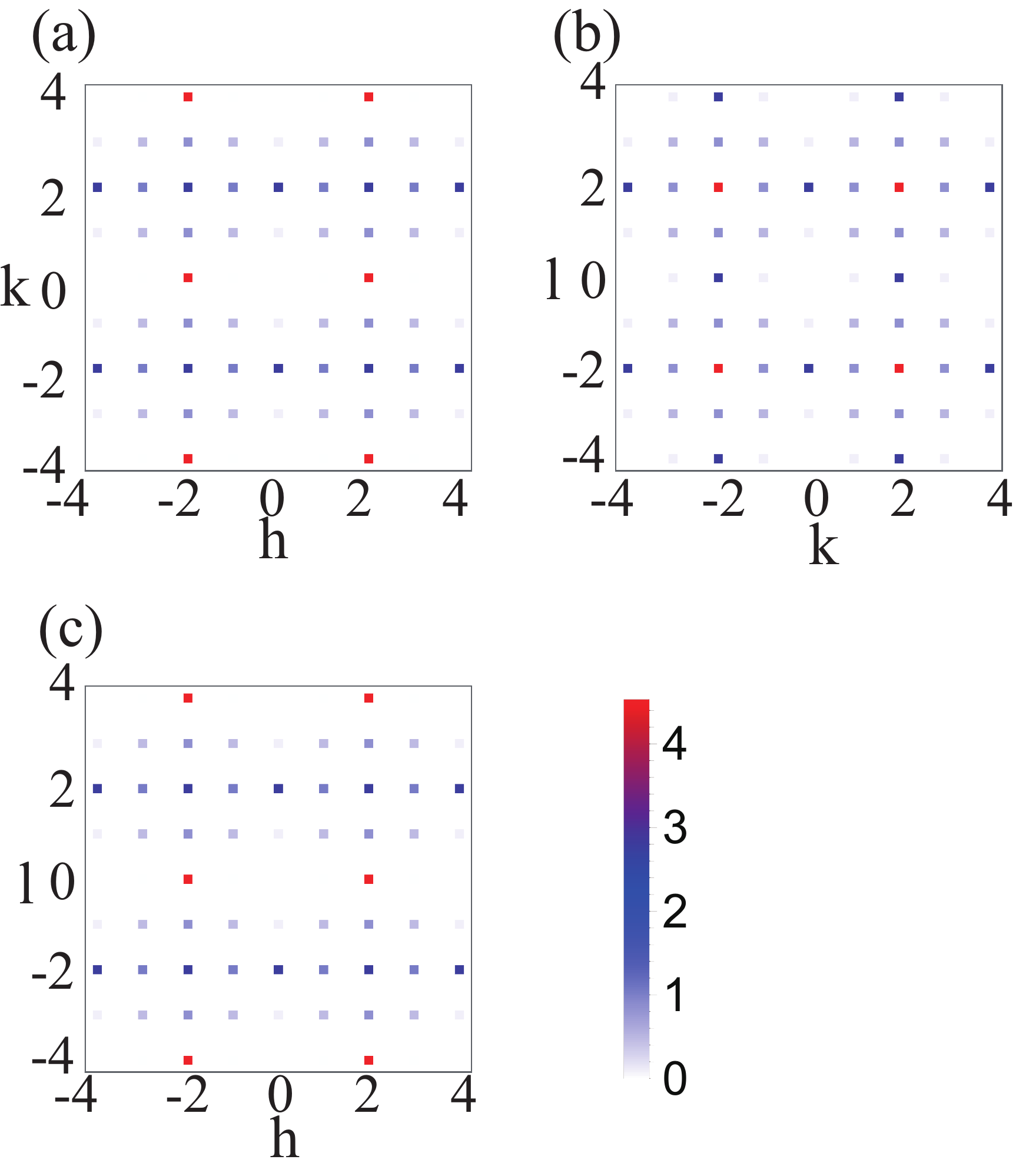}
\vspace{-10pt}
\caption{
(Color online) Spin structure factor of the $\mathbb{Z}_6^{\mathrm{2p}}$-$yz$ state on the (a) $(hk0)$, (b) $(kl0)$ and (c) $(h0l)$ planes.
} 
\label{fig:SF_Z62p} 
\end{center}
\end{figure}

\begin{figure}[!htb]
\begin{center}
\includegraphics[width=7cm]{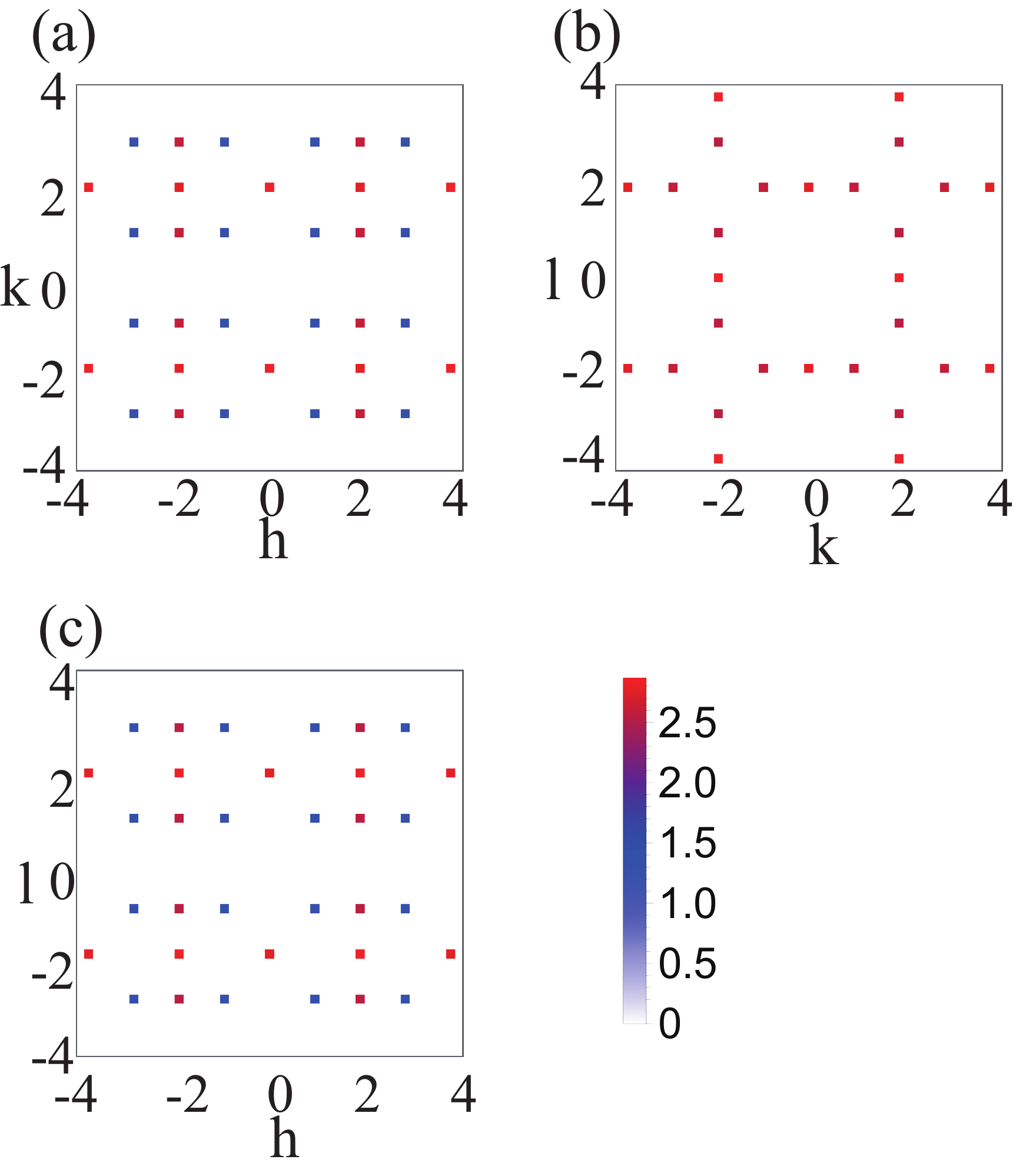}
\vspace{-10pt}
\caption{
(Color online) Spin structure factor of the $\mathbb{Z}_6^{\mathrm{1p}}$-$x$ state on the (a) $(hk0)$, (b) $(kl0)$ and (c) $(h0l)$ planes.
} 
\label{fig:SF_Z61p} 
\end{center}
\end{figure}

Here we provide the static spin structure factors for the $q=0$ magnetic orders.
These structure factors are computed for the long-range ordered cases.
For each $q=0$ order, the structure factor is calculated with the formula
\begin{align}
S(\bm{q}) 
= & \frac{1}{N}\sum_{i,j}
\bm{S}_i \cdot \bm{S}_j e^{-\mathrm{i}\bm{q}\cdot(\bm{r}_i-\bm{r}_j)}.
\end{align}
Here, $N$ is the number of spin moments and $\bm{r}_i$ represents the real space position of the moment $\bm{S}_i$ at site $i$.
Note that for any $q=0$ state the Fourier component of $\bm{S}_i$ is non-vanishing only when the wave vector $\bm{q}$ is equal to a reciprocal lattice vector.
Accordingly, the structure factor has nonzero peaks only at the reciprocal lattice vectors.
The structure factors for the $q=0$ orders are plotted in Figs. \ref{fig:SF_Z2}, \ref{fig:SF_Z62p} and \ref{fig:SF_Z61p}, where $\bm{q} = 2 \pi (h,k,l)$, with $h,k,l$ being integers.
In these figures, one can see that the structure factor is zero at $\bm{q}=0$ for all three states.\cite{SF_footnote}
It is also seen that for the $\mathbb{Z}_2$ windmill state,
the structure factor patterns for all three planes are the same.
This results from the $C_3$ rotational invariance of the $\mathbb{Z}_2$ state.
On the other hand,
for the $\mathbb{Z}_6^{\mathrm{2p}}$-$yz$ and $\mathbb{Z}_6^{\mathrm{1p}}$-$x$ states, the structure factor pattern on the $(0kl)$ plane is different from the others, because of the broken $C_3$ rotational symmetry.
Moreover, $\mathbb{Z}_6^{\mathrm{2p}}$-$yz$ and $\mathbb{Z}_6^{\mathrm{1p}}$-$x$ states show different patterns in the structure factor. 
These differences may be used to distinguish three different $\bm{q}=0$ states.

\section{Estimation of the energy barrier by the single spin rotation process
\label{app:energy_barrier}}
\begin{figure*}[!htb]
\begin{center}
\includegraphics[width=16cm]{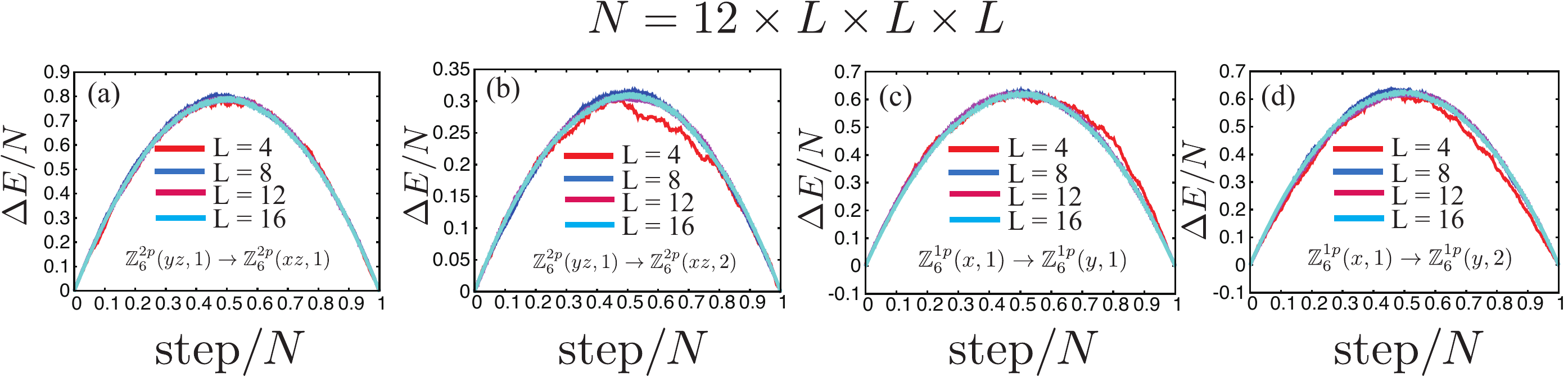}
\vspace{-10pt}
\caption{
(Color online) 
The energy barrier
as a function of the number of steps of single-spin rotations.
The initial and final states 
are described
in each graph.
} 
\label{fig:eb} 
\end{center}
\end{figure*}

In this appendix,
we provide a quantitative estimation of the energy barrier 
between members of the six degenerate
$\mathbb{Z}_6^{2p}$/$\mathbb{Z}_6^{1p}$ states.
The direct path to move from one member to another 
in the spin configuration space is 
obtained by rotating the spins one-by-one 
until the spin-configuration reaches another member. 
In the estimation of the energy barrier, 
we rotate spins in random order
and compute the energy per site measured from the ground-state energy
as a function of the number of spin rotations.

Figure \ref{fig:eb} shows the energy barrier for 
several pairs of the $\mathbb{Z}_6$ states
as a function of the number of spin rotations
for various system sizes.
We can see 
that the energy barrier has a peak at 
$\mathrm{step}/N \sim 0.5$
(i.e., when half of the spins are rotated).
We can also see that the curves for different system sizes are scaled into a single curve.
This indicates that the energy barrier is proportional to the system size. 
Therefore, if a short-range correlation is formed in a relatively large region,
the energy cost to overcome the barrier is very large.
Notice that this energy barrier arises due to the fact that 
the spin configuration is taken out of the ground-state manifold of the 
unperturbed Heisenberg model
during the spin rotation processes.



\begin{thebibliography}{1}
\bibitem{SF_footnote}
It is due to the fact that the $q=0$ states do not have net magnetization.
To be specific, we have 
$S({\bf q}=0) =\sum_{i,j} \bm{S}_i(0) \cdot \bm{S}_j(0)
= \sum_i  \bm{S}_i(0)   \cdot \sum_j \bm{S}_j(0)  =0$
since $ \sum_i \bm{S}_i(0) =0$.
\end{thebibliography}
\end{document}